\documentclass[12pt,preprint,natbib]{aastex}

\usepackage{longtable}
\usepackage{epstopdf}  
\usepackage{ifthen}
\usepackage{rotating}

\singlespace

\newcounter{fignum}

\begin{document}
\bibliographystyle{aj}

\setlength{\footskip}{0pt} 

\begin{center}
  \vspace*{\fill}
  \textbf{The Increasing Rotation Period of Comet 10P/Tempel 2} \\

 \vfill
Matthew M. Knight\footnote{Lowell Observatory, 1400 W. Mars Hill Rd, Flagstaff, AZ 86001}, Tony L. Farnham\footnote{Department of Astronomy, University of Maryland, College Park, MD 20742-2421}, David G. Schleicher$^{1}$, Edward W. Schwieterman\footnote{Physics and Space Sciences, Florida Institute of Technology, 150 W. University Blvd, Melbourne, Florida 32901}\\
  \bigskip
  Contacting author: knight@lowell.edu\\
  \bigskip
  \bigskip
  \bigskip
  \textit{The Astronomical Journal}\\
  Accepted September 13, 2010\\
  \bigskip
  \bigskip
  \vfill
  Manuscript pages: 22 pages text\\
  3 tables \\
  6 figures \\
  \vfill
\end{center}


\newpage
\section*{Abstract}
We imaged comet 10P/Tempel 2 on 32 nights from 1999 April through 2000 March. R-band lightcurves were obtained on 11 of these nights from 1999 April through 1999 June, prior to both the onset of significant coma activity and perihelion. Phasing of the data yields a double-peaked lightcurve and indicates a nucleus rotational period of 8.941 $\pm$ 0.002 hr with a peak-to-peak amplitude of $\sim$0.75 mag. Our data are sufficient to rule out all other possible double-peaked solutions as well as the single- and triple- peaked solutions. This rotation period agrees with one of five possible solutions found in post-perihelion data from 1994 by Mueller and Ferrin (1996, Icarus, 123, 463--477), and unambiguously eliminates their remaining four solutions. We applied our same techniques to published lightcurves from 1988 which were obtained at an equivalent orbital position and viewing geometry as in 1999. We found a rotation period of 8.932 $\pm$ 0.001 hr in 1988, consistent with the findings of previous authors and incompatible with our 1999 solution. This reveals that Tempel 2 spun-down by $\sim$32 s between 1988 and 1999 (two intervening perihelion passages). If the spin-down is due to a systematic torque, then the rotation period prior to perihelion during the 2010 apparition is expected to be an additional 32 s longer than in 1999.


\begin{description}
\item{\textbf{Keywords}:} comets: general -- comets: individual (10P/Tempel 2) -- methods: data analysis -- techniques: photometric
\end{description}

\newpage

\section{Introduction}
\label{intro}
Comet 10P/Tempel 2 was discovered by E.W.L. Tempel on 1873 July 4. It was recovered in 1878, but not during the 1883 or 1889 apparitions. It has been observed on every return since 1894 except three (1910, 1935, and 1941) when it was particularly poorly placed for observing \citep{kronk03, kronk07, kronk09}. While telescopic improvements now allow it to be observed at every apparition, the roughly 5.5 year period results in apparitions which alternate between favorable and unfavorable viewing geometries. Consequently, Tempel 2 was well placed for observing in 1978, 1988, and 1999, but poorly placed in 1983, 1994, and 2004 when it reached perihelion on the far side of the Sun. 

Tempel 2 was extensively observed during its favorable 1988 return. Because it is only weakly active until shortly before perihelion (\citet{sekanina79} and references therein), a nucleus lightcurve can be measured. This allowed a number of investigators to measure the rotation period. The earliest results came from \citet{jewitt88} who found likely periods of 8.9 $\pm$ 0.1 hr or 7.5 $\pm$ 0.1 hr. \citet{jewitt89} extended these observations to near perihelion, concluding that the rotation period was 8.95 $\pm$ 0.01 hr. Comparable periods were determined by \citet{ahearn89} (8.9 hr) and \citet{wisniewski90} (8.93 hr).
\citet{sekanina91} combined these data to determine a sidereal period of 8.93200 $\pm$ 0.00006 hr.

\citet{mueller96} observed Tempel 2 for three nights 7--9 months after perihelion in 1994, after coma activity had subsided.  Their rotation coverage was insufficient to determine a unique period solution, and they found five possible periods due to aliasing: 8.877 hr, 8.908 hr, 8.939 hr, 8.971 hr, and 9.002 hr.
All of these periods were incompatible with the 1988 data, and they concluded that ``the period is definitely different between the 1988 and the 1994 apparitions.'' 
Because of aliasing, it was not known if Tempel 2 had spun-up or spun-down since 1988, merely that its rotation period had changed.

The idea of comet rotation periods changing due to outgassing dates to \citet{whipple50} and his icy conglomerate model of the nucleus, with various authors since then arguing for spin-up or spin-down (see \citet{samarasinha04} for a thorough review). Numerical modeling has shown that, depending on the initial conditions, either spin-up or spin-down is possible, with sustained outgassing over many orbits leading to spin-up and ultimately nucleus splitting \citep{samarasinha95,neishtadt02,gutierrez03}. Measurement of a change in the rotation period of a comet, combined with detailed knowledge of the spin axis orientation, nucleus size and shape, and the outgassing rate can constrain the bulk density and internal structure. Due to the difficulty of studying comet nuclei directly, these properties are only well known for a handful of comets, mostly the result of spacecraft visits.

While simulations have shown that changes in rotation period should be common, a strong case can be made for a changing rotation period in only a few comets: 10P/Tempel 2 (sign of the change unknown prior to this work; \citet{mueller96}), 2P/Encke (spin-up; \citet{fernandez05}), 9P/Tempel 1 (spin-up; \citet{belton07, chesley10}), and Comet Levy (1990c = C/1990 K1; spin-up; \citet{schleicher91,feldman92a}). The paucity of clear detections of this phenomenon is probably due to the difficulty of obtaining high quality datasets of the same comet over multiple apparitions. By combining our extensive data obtained during the 1999 apparition with those of previous authors from 1987--1988 and 1994, this work will show that 10P/Tempel 2 is the first comet known to spin-down.

We observed Tempel 2 from 1999 April until 2000 March, obtaining images in broadband and narrowband optical wavelengths. Observations prior to perihelion (1999 September 8) were primarily obtained with a broadband R filter to measure the nucleus lightcurve. Once activity began in earnest, observations were primarily obtained with narrowband comet filters \citep{farnham00} to study coma morphology. In this paper, we consider only the pre-perihelion nucleus lightcurve in order to resolve the ambiguity about the sign of the change in rotation period and to conclusively demonstrate that Tempel 2 has spun-down since 1988. A second paper (Paper 2) is planned and will utilize data obtained during 2010 in concert with data obtained during the active phase of the 1999 apparition to determine the pole orientation and the location of any active areas.

The layout of the paper is as follows. We summarize our observing campaign and data reductions in Section~\ref{observations}. In Section~\ref{modeling} we analyze the data, determine the rotation period, compare it with earlier datasets, and consider the effects of viewing geometry, coma contamination, and the lightcurve asymmetry. Finally, in Section~\ref{discussion} we discuss the implications of our results and make predictions for the 2010 apparition.

\section{Observations and Reductions}
\label{observations}
\subsection{Observing Overview}
We obtained images of Tempel 2 on a total of 32 nights between 1999 April and 2000 March, with sampling at monthly or shorter intervals. For the study presented here, we use only the data obtained from 1999 April through 1999 June, with the dates and observing circumstances listed in Table~\ref{t:obs_summary}. Additional pre-perihelion data were obtained on 1999 July 10, July 16, August 4, August 5, and September 2. However, these nights all had poor weather and the data were unusable for the present study. By 1999 October, the observing window was very short and the comet had extensive coma contamination, making a period determination challenging.


The April and May observations were obtained at the Lowell Observatory Perkins 1.8-m telescope with the SITe 2K CCD. On-chip 4$\times$4 binning of the images produced a pixel scale of 0.61 arcsec. The June observations were obtained at the Hall 1.1-m telescope with the TI 800 CCD. On-chip 2$\times$2 binning produced images with a pixel scale of 0.71 arcsec. At various times during these runs, we used broadband Kron-Cousins V and R filters and the HB narrowband comet filters \citep{farnham00}. However, as planned, only the R filter measurements are extensive enough for the lightcurve analysis discussed here, so we limit this study to those data. Comet images were guided at the comet's rate of motion. The comet showed evidence of a coma throughout the apparition, progressively increasing with time (see Section~\ref{coma_removal}).

\subsection{Reductions}
The data were reduced using standard bias and flat field techniques. Landolt standard stars \citep{landolt92} were observed to determine the instrumental magnitude and extinction coefficients on 1999 June 8--11. We observed HB narrowband standard stars \citep{farnham00} on all photometric nights, and used these stars as a bootstrap to estimate absolute R-band calibrations for the photometric nights on which Landolt standard stars were not observed.

Fluxes were extracted by centroiding on the nucleus and integrating inside circular apertures, with the median sky calculated in an annulus centered on the nucleus with inner and outer radii $\sim$33 and $\sim$40 arcsec, respectively. 
By extracting fluxes through a series of circular apertures (3, 6, 9,...30 arcsec radius), we monitored the lightcurve for incursion from passing stars which show up in the larger apertures earlier than the smaller apertures. The 6 arcsec radius aperture gave the most coherent lightcurve and was used for photometric analysis; this was selected to be large enough to include most of the light from the nucleus even when the seeing was poor and the nucleus PSF was large, while avoiding as much contamination from passing stars as possible. 


In order to produce usable lightcurves from non-photometric nights, we applied extinction corrections and the absolute calibrations from photometric nights with the same telescope and instrument configuration. Thus, we applied the absolute calibrations from April 19 to April 17--18 and May 26--27, and the calibrations from June 23 to June 22. The deviation of the brightness from photometric nights helps give an estimate of how much obscuration was affecting the non-photometric nights. Field stars on each image were monitored on non-photometric nights to adjust the comet's magnitude for varying obscuration during the night (discussed in the following subsection).

\subsection{Comparison Star Corrections}
\label{comp_stars}
Relative photometry of the comet with respect to field stars was carried out. However, not enough of the same field stars were available during the entire night and we therefore replaced stars which left the field of view with new ones as they entered. When possible we used only stars brighter than m$_R$ = 15, but if fewer than three were available, we used fainter stars, going as faint as m$_R$ = 16. Typically between three and six comparison stars were available at a given time.

Reliable catalog magnitudes were not available for enough comparison stars to use comparison stars for absolute calibrations. Therefore, we first used the median of the seven brightest measurements for a given star during the night as its least obscured brightness and determined the adjustment necessary to bring all fainter measurements into agreement. The comparison star correction for an image was the median offset from each star's least obscured brightness during the night for all comparison stars on a given image. Magnitude corrections derived from comparison stars were applied on the nights specified as non-photometric in Table~\ref{t:obs_summary}. No comparison star corrections were applied for the photometric nights after first using the comparison stars to confirm that the nights were indeed photometric. Nightly median comparison star corrections were 0.1--0.4 mag, although some corrections exceeded 1.0 mag on 1999 April 18.

Since the least obscured brightness of the field stars will be fainter than their ideal brightness if it had been photometric, this technique will introduce a systematic nightly magnitude offset from night to night. Therefore, after correcting the relative photometry with field stars, the entire lightcurve for each night was adjusted so that the peaks for all nights of data were at the same magnitude ($\Delta$m$_2$, discussed in Section~\ref{mag_adjustments}).
If the conditions varied during a night such that there was less obscuration while certain stars were observed but more obscuration while other stars were observed, the uncertainty in the comparison star correction will increase. Hence, the lightcurves on non-photometric nights may exhibit more scatter.

\begin{figure}
  \centering
  \epsscale{0.65}
  \plotone{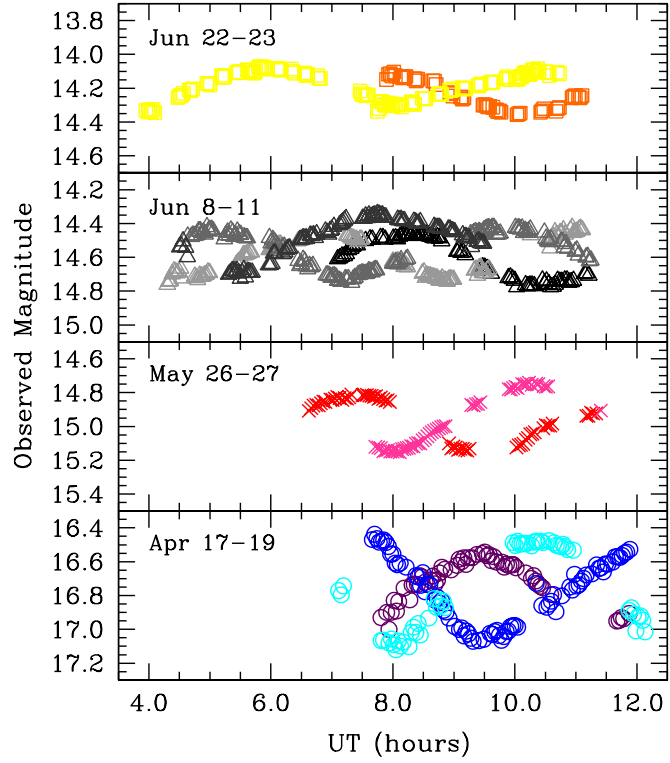}
  \caption[Unphased 1999 data]{Observed R-band magnitudes (m$_R$) for our 1999 data plotted as a function of UT on each night. The magnitudes have had absolute calibrations, extinction corrections, and comparison star corrections applied and are given in column (3) of Table~\ref{t:photometry}. Starting with the bottom panel and moving up, the first panel (circles) displays April 17 (purple), April 18 (blue), and April 19 (cyan). The second panel (crosses) displays May 26 (red) and May 27 (pink). The third panel (triangles) displays June 8 (black), June 9 (light gray), June 10 (medium gray), and June 11 (dark gray). The top panel (squares) displays June 22 (orange) and June 23 (yellow). Note that throughout the time of interest, decreasing $r$ and $\Delta$ caused the comet to brighten while at the same time increasing coma damped out the amplitude variations of the nucleus. Thus the range of magnitudes is different in each panel, although the vertical scale is held constant at 1 mag to emphasize the decreased amplitude.}
  \label{fig:orig_lcs}
\end{figure}

The calibrated, extinction corrected, and comparison star corrected  R magnitudes (m$_R$) are plotted in Figure~\ref{fig:orig_lcs} as a function of UT. This indicates how much nightly coverage was obtained. In April and May, Tempel 2 was observed for about five hours, while in June it was observed for about seven hours, with the observing window moving earlier each night. Because the comet brightened during the apparition and the coma increased, the median magnitude is different in each figure and later nights have smaller amplitudes (the vertical scale is held fixed in all panels but the range of magnitudes varies from panel to panel). In order to create a uniform dataset for period analysis and to study the lightcurve amplitudes, we removed the coma, corrected for changing $r$, $\Delta$, and $\phi$, and adjusted for nightly offsets. These adjustments will be discussed in the next two sections.

\begin{figure}
  \centering
  \epsscale{0.55}
  \plotone{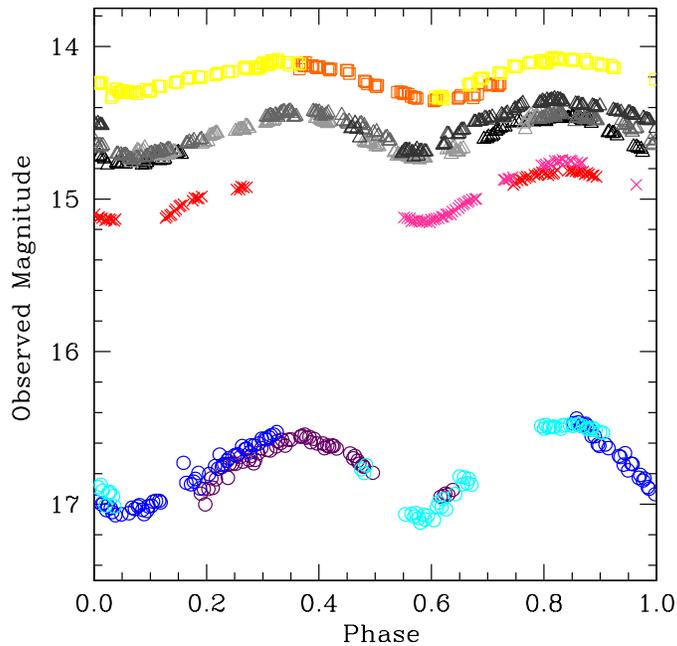}
  \caption[Unphased 1999 data]{Observed R-band magnitudes (m$_R$) for our 1999 data phased to our best period of 8.941 hr (discussed in Section~\ref{1999_period}). The magnitudes have had absolute calibrations, extinction corrections, and comparison star corrections applied and are given in column (3) of Table~\ref{t:photometry}. Zero phase was set at perihelion (1999 September 8.424). The points are as given in Figure~\ref{fig:orig_lcs}. Note that throughout the time of interest, decreasing $r$ and $\Delta$ caused the comet to brighten while at the same time increasing coma damped out the amplitude variations of the nucleus. This figure illustrates the phase coverage obtained during each run, and shows that the lightcurves could be phased even without removing coma contamination.}
  \label{fig:orig_lcs_phased}
\end{figure}

In Figure~\ref{fig:orig_lcs_phased} we show the same data as in Figure~\ref{fig:orig_lcs} but phased to 8.941 hr (our best period which will be discussed in Section~\ref{1999_period}). This figure emphasizes the increase in brightness and decrease in amplitude throughout the run, as later nights are higher in the figure and have smaller amplitudes. It also demonstrates the phase coverage obtained during each run, making it possible to determine the rotation period even without removing the coma contamination.


\subsection{Removal of Coma Contamination}
\label{coma_removal}
\begin{figure}
  \centering
  \epsscale{0.55}
  \plotone{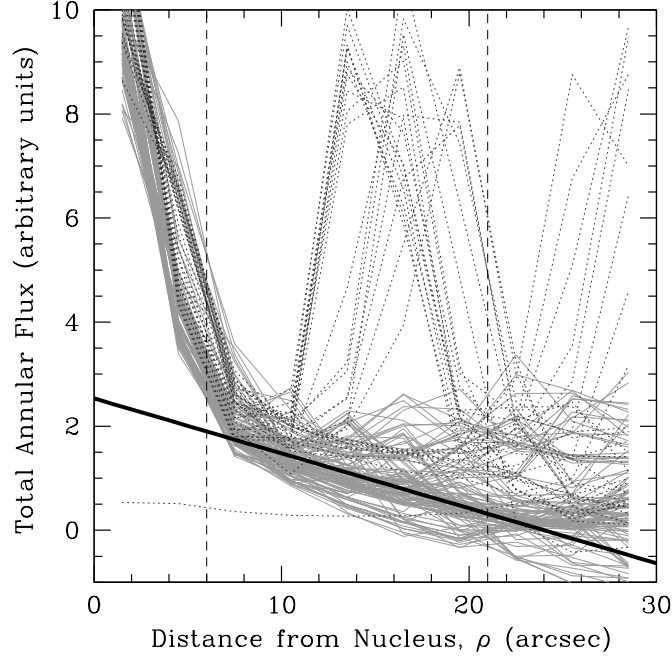}
  \caption[Coma removal on 1999 June 9]{Representative nightly radial profile (1999 June 9). Each curve is composed of the total annular flux (in arbitrary units) in 3 arcsec wide annuli. The solid light gray curves are the radial profiles for images which were used to estimate a nightly median coma profile. The dotted dark gray curves are the radial profiles for images which had contamination by background stars and were not used to determine the nightly median coma profile. The heavy black curve is the nightly median coma profile. This was determined by taking the median of all the light gray curves at the midpoint of each annulus, then fitting a straight line to this nightly radial profile for annuli between 6--21 arcsec (the vertical dashed lines), and extrapolating the fit in to the nucleus and out to 30 arcsec. If the coma decreased as $\rho$$^{-1}$ the coma profile would be a horizontal line. The coma profile was never a horizontal line due to the wings of the nucleus PSF, contamination from background stars, and possibly a non-$\rho$$^{-1}$ fall off of the coma. We fit a straight line as a first order approximation. Higher order fits varied more from image to image. See the text for further discussion.}
  \label{fig:coma_removal}
\end{figure}

While Tempel 2 is not strongly active, some coma was visible throughout the apparition. The coma contamination is, in general, not large and the rotation period can be determined without coma removal. However, to ensure that it was not affecting our results and to compare the amplitude of the lightcurve with those obtained by other authors, we removed the coma inside the monitoring aperture. 
For unchanging grains flowing radially outward from the nucleus at a constant velocity, the coma flux per pixel decreases as $\rho$$^{-1}$, where $\rho$ is the projected distance from the nucleus. Since the area of equally spaced annuli increases as $\rho$, the total coma flux in each annulus should be constant. In reality, the coma often does not fall off as $\rho$$^{-1}$ and there are factors which may cause further deviation from a constant total flux per annulus such as contamination by background stars or cosmic rays, wings of the nucleus PSF, and imperfect background removal. However, we found that a linear fit to the total annular flux as a function of annular distance from the nucleus ($\rho$) provides a reasonable first order approximation of the coma. This is illustrated in Figure~\ref{fig:coma_removal} using 1999 June 9 as a representative night.


We calculated the total flux in 3 arcsec wide annuli centered on the nucleus (e.g., 0--3 arcsec, 3--6 arcsec,...27--30 arcsec) to create radial profiles (total annular flux as a function of $\rho$) for each image. These are plotted as solid light gray curves (images without significant contamination from background stars) and dotted dark gray curves (images with significant contamination from background stars) in Figure~\ref{fig:coma_removal}. We then computed the median total flux in each annulus for the night, ignoring images with obvious contamination from background stars. Next, we fit a straight line to the total annular flux as a function of distance from the nucleus, $\rho$, (the heavy black line in Figure~\ref{fig:coma_removal}) from $\rho$ = 7.5--19.5 arcsec for each image (where $\rho$ = 7.5 arcsec was the center of the 6--9 arcsec annulus and $\rho$ = 19.5 arcsec was the center of the 18--21 arcsec annulus). This range was chosen to exclude as much of the signal from the nucleus ($\rho <$ 6 arcsec) as possible and to minimize contamination from passing stars ($\rho >$ 21 arcsec). We extrapolated the fit in to the nucleus ($\rho$ = 0 arcsec) and out to $\rho$ = 30 arcsec. The total coma annular flux was removed from the total annular flux to give the coma corrected total annular flux (i.e. the total nucleus annular flux), which was integrated and converted back to magnitudes. Since the monitoring aperture was 6 arcsec in radius, the total coma which was removed for a night was the sum of the total coma annular flux in the 0--3 and 3--6 arcsec radius annuli.

When determining the nightly coma, we excluded images with stars which obviously altered the fit. Stars that contaminate the larger annuli tend to result in an under-removal of coma in the monitoring aperture, while stars closer to the nucleus tend to result in an over-removal of the coma in the monitoring aperture. While we excluded the images with obvious contamination from background stars, fainter stars undoubtedly remain. Since more background stars pass through the larger annuli than the smaller annuli, this results in a systematic under-removal of the coma.

As shown in Figure~\ref{fig:coma_removal}, typical profiles appear to have some curvature. This implies that more coma should be removed than is accomplished with a linear fit.  A significant contributor to the curvature at small $\rho$ is likely the wings of the nucleus PSF, while at large $\rho$, the coma signal may simply be too small and is swamped by uncertainty in the background removal. We considered higher order fits to better match the curvature; a quadratic fit removed 20--60\% more coma while an exponential fit was poor as it often implied negative nucleus counts. Although these fits removed more coma than the linear fit, they varied more from image to image, resulting in wider variance in the estimated coma. Also, small fluctuations in the larger annuli, where there is very little coma, have a large effect on the estimate of the coma contamination for these fits. Therefore, we concluded that a simple linear fit provided the best compromise for approximating the coma profile. We note that if a different fit to the coma were used, it would somewhat alter the amplitude of the lightcurve but would not change the location of its extrema. Thus, the coma removal technique does not affect the key result of this paper, the period determination (Section~\ref{1999_period}).

To test that our use of a single nightly coma correction was appropriate, we determined linear coma fits for each usable image on a night. We saw no systematic variation in the estimated coma flux within the photometric aperture as a function of rotational phase throughout the entire campaign, confirming that the use of a single nightly median correction was reasonable. As another test of the coma removal technique, we investigated the total annular nucleus flux remaining after coma removal. The median fraction for all nights of the nucleus flux contained in the 3 arcsec radius aperture relative to the 9 arcsec radius aperture was 85\%. The median for the 6 arcsec radius aperture relative to the 9 arcsec radius aperture was 99\%. Apertures larger than 9 arcsec in radius were within $\pm$1\% of the 9 arcsec radius aperture flux before deviating at $\rho$ $>$ 21 arcsec (which was beyond the range of the coma fit). These ratios were relatively constant throughout the apparition. The ratio of the 3 arcsec radius aperture relative to the 9 arcsec radius aperture varied with seeing changes, but the 6 arcsec radius aperture did not change appreciably. This confirms that the 6 arcsec radius aperture is appropriate for photometric monitoring, and that the light lost by not going to larger apertures should be minimal and roughly the same fraction in all images, causing no effect on the amplitude of the lightcurve, regardless of seeing. 

\subsection{Magnitude Adjustments}
\label{mag_adjustments}
The viewing circumstances changed significantly during our observations. Therefore, we adjusted the nucleus magnitudes using the standard asteroidal normalization
\begin{equation}
\label{e:mag_norm}
m_{R}(1,1,0) = m_{R,cr} - 5log(r\Delta) - \beta \alpha
\end{equation}
where m$_R$(1,1,0) is the normalized magnitude at $r$ = $\Delta$ = 1 AU and $\alpha$ = 0$^\circ$, m$_{R,cr}$ is the apparent magnitude, m$_R$ (which has had the absolute calibrations, extinction corrections, and comparison star corrections applied), with the coma removed, $r$ is the heliocentric distance (in AU), $\Delta$ is the geocentric distance (in AU), $\beta$ is the linear phase coefficient, and $\alpha$ is the phase angle. $\beta$ is typically 0.03--0.04 mag deg$^{-1}$ for comets, and we used $\beta$ = 0.032 mag deg$^{-1}$ since it minimized the $\Delta$m$_2$ adjustment (discussed in the following paragraph). Equation~\ref{e:mag_norm} removes the secular variation in brightness and allows comparison of all lightcurves on a similar scale. The geometric corrections are given as $\Delta$m$_1$ in column (11) in Table~\ref{t:obs_summary} at the midpoint of each night's observations.
 
The geometric corrections do not always bring the lightcurves from different nights to the same peak brightness, with variations as high as 0.33 mag. This is due to a number of factors, including: bootstrapping Landolt standard stars from HB standards on photometric nights when Landolt stars were not observed; the application of absolute calibrations on non-photometric nights; comparison stars that are normalized to their brightest point in the night rather than a catalog value; and the shape of the coma removed.
Therefore, we introduced an additional adjustment, $\Delta$m$_2$ (column (12) in Table~\ref{t:obs_summary}), to adjust the individual lightcurves to a common peak brightness. It should be noted that this factor is introduced to simplify the rotation period analysis, and should not be interpreted as representing any particular physical property.


After phasing the data with preliminary $\Delta$m$_2$ values, we refined $\Delta$m$_2$ so that all lightcurves were aligned at the peak near $\sim$0.85 phase when possible. April 17 and June 22 were aligned using the peak near $\sim$0.35 phase. Since April 18 and June 9 did not have conclusive peaks, $\Delta$m$_2$ for these nights was estimated based on the cluster of points near phase 0.85.

\subsection{Data}
The photometry is given in Table~\ref{t:photometry}. Columns (1) and (2) are the UT date and time (at the telescope) at the midpoint of each exposure (adjustments for light travel time are given in Table~\ref{t:obs_summary}). Column (3) is m$_R$, the observed R-band magnitude after photometric calibrations, extinction corrections, and comparison star corrections have been applied. Column (4) is m$_R$$^*$, the coma-removed, reduced magnitude m$_R$(1,1,0) corrected by $\Delta$m$_2$ so that all nights have a similar peak magnitude. We obtained 1016 data points, of which 124 were discarded due to contamination from background stars, tracking problems, or cosmic ray hits. 


While there are several adjustments which have been applied to arrive at the m$_R$$^*$ values listed in Table~\ref{t:photometry}, they do not affect the primary focus of the paper: the rotation period determination. The corrections for geometry ($\Delta$m$_1$) and nightly offsets ($\Delta$m$_2$) shift individual nightly lightcurves up or down, but the locations of the extrema in rotational phase do not change. While the choice of coma removal algorithm affects the amplitude of the lightcurve, it also leaves the phase of the extrema unchanged. Therefore, the same rotation period could be determined directly from the m$_R$ values prior to the corrections.

Due to the large number of sources of error for each data point (photon uncertainty in comet and background flux, extinction correction, comparison star correction, coma removal), we estimated the effective uncertainty in the magnitude by fitting a smoothed spline through each night's lightcurve and subtracting the spline fit. The uncertainty for the night was estimated as the standard deviation of the residuals. The uncertainty was not always constant throughout a night since seeing conditions and obscuration varied on the non-photometric nights, so on some nights we estimated the uncertainty for subsections of the lightcurve. Variations in the comparison star magnitudes were used as an additional indicator of when the uncertainty changed during a night. The range of uncertainties for a night are given in column (11) in Table~\ref{t:obs_summary}.

\section{Modeling and Interpretation}
\label{modeling}
\subsection{Determining the Rotation Period in 1999}
\label{1999_period}
We defined zero phase to be perihelion (1999 September 8.424), and accounted for the light travel time (column (8) in Table~\ref{t:obs_summary}) before phasing the data. We used an interactive period search routine within the Data Desk data analysis package\footnote{http://www.datadesk.com.} which updates the phased lightcurves on the fly, allowing us to easily scan through potential periods.
We looked for alignment of extrema since they can be identified in phase space even if the magnitudes or amplitudes are different. As the correct solution is approached, the extrema tend to be systematically shifted in phase based on the (color-coded) date of observation, making it easy to quickly hone in on the optimal solution. However, it is difficult to quantitatively estimate the uncertainty in the period since the alignment of the features is probably well away from the optimal solution by the time the eye can distinguish it, therefore overestimating the uncertainty.

We applied this technique to the m$_R$$^*$ data given in column (4) of Table~\ref{t:photometry}, finding an observed, i.e. synodic, period of 8.941 $\pm$ 0.002 hr for the combined dataset (April--June). We examined all period solutions between 4.47 hr (the single-peaked solution) and 13.41 hr (the triple-peaked solution) and 8.941 hr (the double-peaked solution) is the only viable solution between these extremes.
The double-peaked shape is expected for a triaxial ellipsoid and was observed by previous authors in 1988 and 1994. We plot the m$_R$$^*$ data phased to 8.941 hr in the top panel of Figure~\ref{fig:1999_data}. The bottom panel is phased to 8.932 hr (the period solution from 1988, discussed in the following subsection).


The uncertainty in the period was estimated from the smallest offset that produced lightcurves that were visibly out of phase. Figure~\ref{fig:best_period} illustrates an exaggeration of this process using three periods: 8.937 hr (top), 8.941 (middle) and 8.945 (bottom). Note that these periods are separated by double the estimated error to emphasize the trend in the location of the extrema. We plot the m$_R$$^*$ data offset vertically by $-$0.015 mag day$^{-1}$ since 1999 April 17 to better show the chronological progression of the lightcurves. In the top panel, later lightcurves (higher on the plot) occur at a later phase, while in the bottom panel later lightcurves occur at an early phase. This clearly illustrates that the correct solution is between the two, making it easy to identify 8.941 hr as the optimal period. The inclusion of the April data is vital for this process, as the larger separation between April and the later datasets decreases the range of possible solutions.


For comparison, we also conducted period searches using Fourier \citep{deeming75} and phase dispersion minimization (PDM; \citet{stellingwerf78}) techniques. The Fourier routine ``Period04''\footnote{http://www.univie.ac.at/tops/Period04/.} \citep{lenz05} found a single-peaked period of 4.4705 $\pm$ 0.0001 hr. Since we expect the lightcurve to be double-peaked, this implies a rotation period twice as long, or 8.9409 $\pm$ 0.0002 hr (the difference in the last digit is due to rounding). The routine determined an uncertainty in frequency and we converted this to a period uncertainty, although we note that the uncertainty of 0.0002 hr seems unrealistically small. We used our own PDM routine, which found a double-peaked period of 8.941 $\pm$ 0.005 hr. We estimated the period uncertainty by finding the range of solutions that had $\theta$ (a measure of the goodness of fit) within 50\% of the minimum $\theta$. The agreement between methods confirms the robustness of our period solution.

\subsection{Different Rotation Periods Due to Geometry?}
\label{geometry}
The default rotation period obtained by simply phasing the data is the synodic period, the length of time until the brightness appears the same from the Earth. This period changes during the apparition because the illuminated portion of the nucleus seen by the Earth varies as both the Earth and the comet move in their orbits. The sidereal rotation period is the rotation period relative to a fixed position in inertial space. A third period is the solar period, which is the time until the Sun is in the same place in the sky as seen by the comet. Because the synodic and solar rotation periods depend on the orientation of the rotational pole, converting them to a sidereal period requires a pole solution.

Since the synodic period also depends on the position of the Earth, we must consider the geometry of the Earth-comet-Sun system during our observations to confirm that the synodic period changes slowly enough that measuring a single synodic period for the whole 10-week time span is sensible. As seen from the comet, the ecliptic longitude of the Earth (column (9) of Table~\ref{t:obs_summary}) and the ecliptic latitudes of the Sun and Earth all changed by less than 5$^\circ$ during our observations, resulting in minimal changes to the measured rotation period. While the ecliptic longitude of the Sun as seen from the comet (column (10) of Table~\ref{t:obs_summary}) changed by $\sim$30$^\circ$, the net effect on the rotation period was still small due to the likely pole orientation (see discussion below), and obtaining a single (synodic) rotation period for the whole time span was valid. 

To confirm this, we investigated the synodic rotation periods which would be observed at different times for pole solutions within 20$^\circ$ in the comet's orbital coordinate system (obliquity and azimuthal angle) of the pole solution given by \citet{sekanina87b}. This range of solutions was chosen to include the pole solution found by \citet{sekanina91} for the 1988 apparition and a preliminary pole solution for our 1999 coma morphology data (we will explore this in detail in Paper 2). We considered three runs discretely: April 17--19, May 26--27, and June 22--23 (the June 8--11 run gave similar results to the June 22--23 run), and determined the synodic--sidereal offset which would be observed for all pole solutions under consideration. The difference between the synodic and sidereal periods was 0.000 to $+$0.003 hr during the April run, 0.000 to $+$0.002 hr during the May run, and $-$0.001 to $+$0.001 hr during the late-June run. Considering the entire 10-week interval as a single large run (i.e. April 17--June 23), the best synodic period differed from the sidereal period by 0.000 to $+$0.002 hr. Thus, the changing geometry during our observations did not introduce an error in the period determined from the entire run larger than the estimated uncertainty in the synodic period of 0.002 hr (although if the April run was considered by itself, the maximum difference was 0.003 hr). Note that the sign of the synodic--sidereal difference reverses for a pole orientation in the opposite direction, i.e. when the comet has retrograde spin and ``north'' is defined in the opposite hemisphere.

The above example suggests that changes in the synodic period may be detectable during an apparition. We considered subsets of our 1999 data to look for such a change. There was no discernible difference when the rotation period was determined using all the April and May data or all the May and June data (both intervals gave acceptable solutions from 8.940--8.942 hr with a best period of 8.941 hr). We also determined the synodic period between every combination of individual nights from different runs for which a synodic period could be determined. We phased each pair of nights and defined the midpoint between the two nights as the date corresponding to the measured synodic period. This method suggests an increase in the synodic period of $\sim$0.002 hr from early-May to mid-June, although the scatter in the individual synodic measurements is very large and the range of acceptable solutions for each pair is generally at least $\pm$ 0.002 hr. If correct, this requires that the ``north'' pole  be in the opposite hemisphere as was determined by \citet{sekanina87b}, i.e. retrograde rather than prograde rotation.

\subsection{Reanalysis of the 1988 Data}
\label{1988_data}
We reanalyzed the publicly available datasets from 1988 (\citet{jewitt89}, \citet{ahearn89}, and \citet{wisniewski90}), defining zero phase at perihelion (1988 September 16.738). The observing scenarios for these datasets are summarized in Table~\ref{t:obs_summary_1988}. \citet{jewitt89} removed coma from their June data but not the February or April data. \citet{ahearn89} obtained optical and thermal IR data simultaneously on different telescopes, and removed coma from the optical data but not from the thermal IR data, which were shown to be nearly free of coma. \citet{wisniewski90} did not remove coma from his data. The 1987 data from \citet{jewitt88} were not included in the 1988 period determination because they were too far away in time and were at a much different location in the orbit, making it impossible to directly compare the synodic rotation periods.


We normalized the published magnitudes to m$_R$(1,1,0) using $\beta$ = 0.032 mag deg$^{-1}$ as was used with our 1999 data. After this normalization, the optical magnitudes still differed by several tenths of a magnitude between runs and authors. We therefore adjusted all the lightcurves to peak at the same magnitude for the maximum near zero phase by applying nightly $\Delta$m$_2$ values (given in column (12) of Table~\ref{t:obs_summary_1988}). We call these data m$_R$$^*$, and used them for our period search. $\Delta$m$_2$ corrects for nightly differences between datasets due to differences in the extinction correction, absolute calibration, and coma removal (or lack thereof). We do not list a $\Delta$m$_2$ for the \citet{ahearn89} 4845 {\AA} data, as we did not use them when phasing the data since the 4845 {\AA} and 6840 {\AA} data were obtained alternately throughout both nights, and their lightcurve shapes were nearly identical (although the 6840 {\AA} data were brighter).

Visually scanning the 1988 data for a double-peaked solution, as was done for our 1999 data, yielded a period of 8.932 $\pm$ 0.001 hr. This period agrees with the synodic periods determined for the 1988 data by \citet{sekanina91} (8.931 $\pm$ 0.001 hr, given in his Table~5) and \citet{mueller96} (8.933 $\pm$ 0.002)\footnote{We did not see evidence of a 13 min offset of the 1988 April data from \citet{jewitt89} as suggested by \citet{sekanina91}, but note that $\Delta$m$_2$ could mask this effect. \citet{sekanina91} adjusted these data by 13 minutes to derive the quoted synodic period, while \citet{mueller96} omitted these data entirely from their period analysis. We include them with no time adjustment.}.

The viewing geometry in 1988 was nearly identical to the viewing geometry in 1999; the ecliptic latitudes of the Earth and Sun changed minimally and the ecliptic longitudes of the Earth and Sun (given in columns (9) and (10), respectively, in Table~\ref{t:obs_summary_1988}) mimicked the changes in 1999. As with the 1999 data (Section~\ref{1999_period}), we considered discrete runs spanning the range of 1988 observations: February 24--28, April 8--12, May 18--22, and June 19--23, and determined the synodic--sidereal difference during each run for the range of pole solutions. Because of the similar viewing geometry, the synodic--sidereal difference was almost the same as in 1999: $+$0.001 to $+$0.002 hr for February 24--28, $+$0.001 to $+$0.002 hr for April 8--12, 0.000 to $+$0.001 hr for May 18--22, and $-$0.001 to $+$0.001 hr for June 19--23. Considering the entire 1988 time span from February 25 until June 30 as a single large run, the difference in the synodic and sidereal rotation periods was 0.000 to $+$0.002 hr for the range of pole solutions. As with 1999, the synodic--sidereal difference changes sign for a pole orientation in the opposite direction. 

\begin{figure}
  \centering
  \epsscale{0.55}
  \plotone{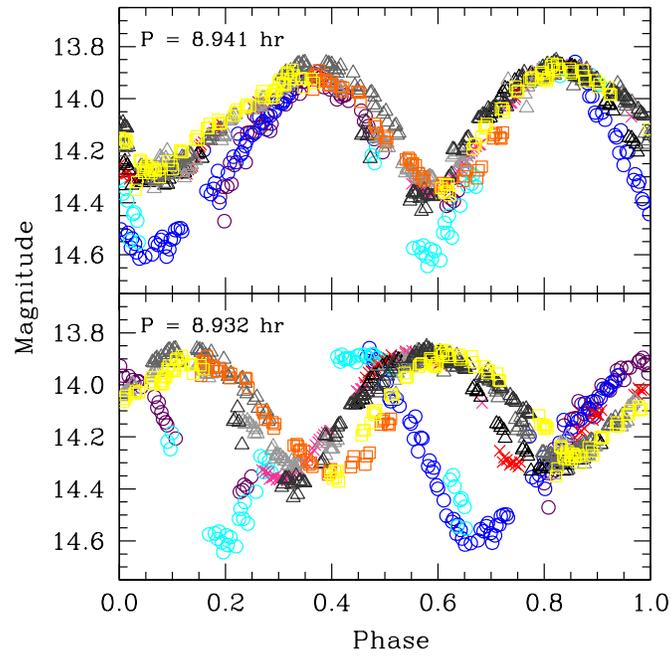}
  \caption[1999 data]{1999 data phased to the 8.941 hour period (top panel) and the 8.932 hr (bottom panel). The magnitudes are m$_R$$^*$ (given in column (4) of Table~\ref{t:photometry}). The midpoint of each image was used and then corrected for the light travel time, with zero phase set at perihelion (1999 September 8.424). The points are as given in Figure~\ref{fig:orig_lcs}.}
  \label{fig:1999_data}
\end{figure}

\begin{figure}
  \centering
  \epsscale{0.55}
  \plotone{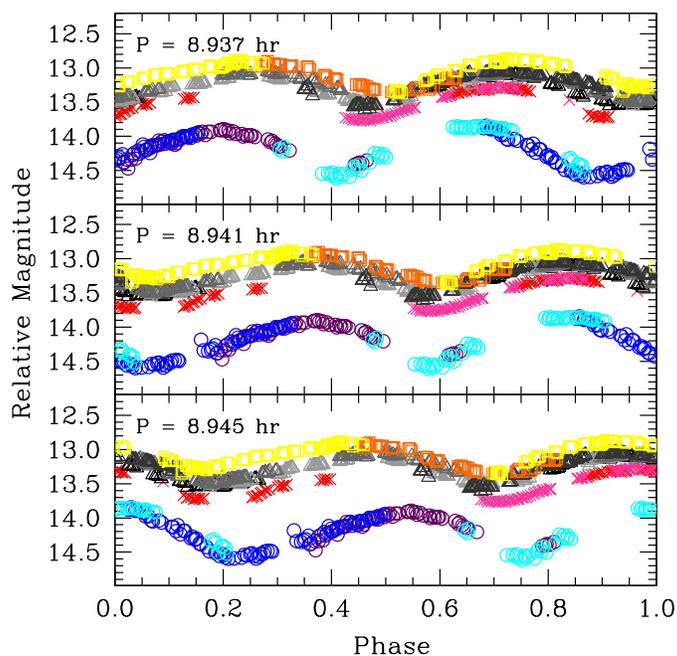}
  \caption[1999 phased data]{Our 1999 data phased to 8.937 hr (top panel), 8.941 hr (middle panel), and 8.945 hr (bottom panel). The magnitudes are m$_R$$^*$ (given in column (4) of Table~\ref{t:photometry}) offset by $-$0.015 magnitudes per day from 1999 April 17. This offset has been applied for display purposes to minimize overlapping data so the trend in the phasing of extrema is more apparent. The midpoint of each image was used and then corrected for the light travel time before phasing, with zero phase set at perihelion (1999 September 8.424). The symbols are as given in Figure~\ref{fig:orig_lcs}. By phasing the data in this manner, we found a best period of 8.941 hr and an uncertainty of $\pm$0.002 hr.}
  \label{fig:best_period}
\end{figure}

\begin{figure}
  \centering
  \epsscale{0.55}
  \plotone{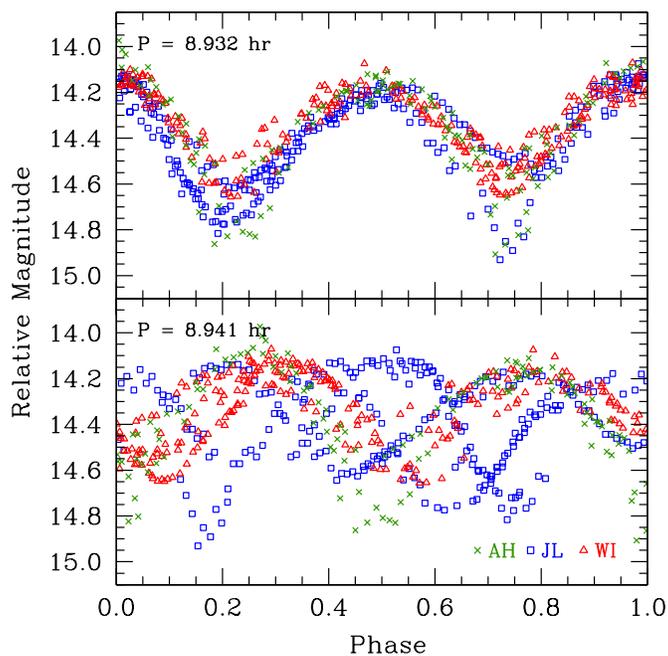}    
  \caption[1988 data]{1988 data from the literature phased to the 8.932 hour period (top panel) and the 8.941 hr (bottom panel). The published data have been normalized to m$_R$(1,1,0) and offset by $\Delta$m$_2$ (given in Table~\ref{t:obs_summary_1988}) to peak at the same magnitude. The times are as given in the literature and corrected for the light travel time, and zero phase was defined at perihelion (1988 September 16.738). The green crosses (AH) are \citet{ahearn89}, the blue squares (JL) are \citet{jewitt89}, and the red triangles (WI) are \citet{wisniewski90}.}
  \label{fig:1988_data}
\end{figure}

Thus we can directly compare our 1988 solution (8.932 hr) with our 1999 solution (8.941 hr). In Figure~\ref{fig:1988_data} we plot the 1988 m$_R$$^*$ data phased to our period solution from 1988 in the top panel and to our period solution from 1999 in the bottom panel. Similarly, we plot our 1999 m$_R$$^*$ data phased to 8.941 hr (top panel) and 8.932 hr (bottom panel) in Figure~\ref{fig:1999_data}. Figures~\ref{fig:1999_data} and \ref{fig:1988_data} demonstrate that the rotation periods were clearly different in 1988 and 1999. Since the synodic periods differed by $\sim$0.009 hr and synodic--sidereal offsets were similar each apparition, we can conclude that the sidereal periods also differed by $\sim$0.009 hr, and thus the rotation period increased from 1988 to 1999.


We looked for a change in the synodic period during the 1988 observations by phasing every combination of pairs of observing runs, e.g. \citet{jewitt89}'s February data with \citet{ahearn89}'s June data, \citet{jewitt89}'s February data with \citet{wisniewski90}'s June data, etc. Assigning the date of each pair as the midpoint between the two runs, there was no discernible trend in the measured synodic period over time, with acceptable values varying from 8.931--8.933 hr with a typical uncertainty of $\pm$0.001 hr.
This is consistent with the conclusions of \citet{jewitt89} who found no change in the rotation period within their uncertainty when analyzing their February, April, and June data as discrete sets. In contrast, \citet{sekanina91} found a decrease in the synodic period during the 1988 observations of $\sim$0.003 hr (his Table 5 and Figure 7). These discrepancies and our possible finding of a slight increase in the rotation period in 1999 (Section~\ref{1999_period}) suggest that \citet{sekanina87b}'s pole solution may not be definitive, with the alternate sense of rotation possible as well (resulting in a ``north'' pole in the exact opposite position). 
We will use the coma morphology visible in our post-perihelion 1999 images, combined with data we will obtain in 2010, to attempt to determine a more robust pole solution in Paper 2.

\subsection{Reanalysis of the 1994 Data}
\label{1994_data}
We also reanalyzed the \citet{mueller96} data from 1994, normalizing the published magnitudes to m$_R$(1,1,0) and visually scanning for possible solutions. The data do not yield an unambiguous double-peaked lightcurve due to undersampling. \citet{mueller96} found five possible solutions: 8.877 hr, 8.908 hr, 8.939 hr, 8.971 hr, and 9.002 hr, and we concur with their results. Only one of these, 8.939 hr, is within the uncertainty of our 1999 solution (8.941 $\pm$ 0.002 hr). 

Using the same range of pole solutions as in Section~\ref{geometry}, we find the maximum difference between the synodic and sidereal periods in October--December 1994 was smaller than $-$0.001 hr (actually about $-$1 s), in agreement with the difference found by \citet{mueller96}. Thus, 
the expected difference between the synodic and sidereal rotation periods is much less than the difference between the five possible solutions. Furthermore, since there were no intervening perihelion passages between the 1994 and 1999 observations, the sidereal rotation period was likely unchanged between these observations. Therefore, we rule out the remaining four possible solutions from \citet{mueller96}, and conclude that Tempel 2 had spun-down by $\sim$0.009 hr ($\sim$ 32 s) between the 1988 and 1994 determinations, which spanned two perihelion passages, and was unchanged between the 1994 post-perihelion observations and our 1999 pre-perihelion observations.

\subsection{Coma Contamination and Lightcurve Amplitude}
\label{amplitude}
Tempel 2 was non-stellar in appearance throughout our 1999 observations. The presence of a coma suppressed the amplitude of the lightcurve, with the suppression increasing later in the apparition as activity increased.  The mean coma removed from the 6 arcsec radius monitoring aperture was 0.20 mag from April 17--19, 0.27 mag from May 26--27, 0.40 mag from June 8--11, and 0.65 mag from June 22--23. We measured lightcurve amplitudes ranging from 0.2--0.5 mag prior to removal of the coma. After coma removal, the April 17--19 data exhibited amplitudes of $\sim$0.75 mag while all subsequent runs had amplitudes of 0.45--0.50 mag. If a quadratic fit to the coma was used instead of a linear fit, the amplitudes increased to $\sim$0.80 in April and 0.50--0.60 in May and June. However, note that as discussed in Section~\ref{coma_removal}, we opted for a more conservative linear fit rather than a higher order fit because the higher order fits varied more from image to image and produced less reliable results. Our measured amplitudes in 1999 are similar to the maximum optical amplitudes reported by other authors in 1988 when the comet was at a similar viewing geometry and activity level: 0.65 $\pm$ 0.05 mag in February and 0.60 $\pm$ 0.05 mag in April (no coma removal; \citet{jewitt89}), 0.5 mag in May (no coma removal; \citet{wisniewski90}), and 0.7 $\pm$ 0.1 mag in June (coma removed; \citet{jewitt89}).

\citet{ahearn89} measured an amplitude of 0.8 mag in their 10.1 $\mu$m data in 1988 June, exceeding all measured optical amplitudes at similar times in both 1988 and 1999. While the amplitudes measured by ourselves and others approach the thermal IR amplitude, all are somewhat lower, and even a quadratic coma removal from our data cannot bring the amplitudes into agreement. This discrepancy raises two possible interpretations. First, the coma removal techniques employed here and by \citet{jewitt89} and \citet{ahearn89} are not adequately estimating the coma inside the photometric aperture. To push the amplitude up to $\sim$0.80, 50--150\% more coma would need to be removed from our May and June data, requiring a significantly steeper curve than even a quadratic fit. Alternatively, the convergence of multiple authors on an optical amplitude of 0.6--0.7 mag in 1988 June and 1999 June may indicate that this is in fact the correct optical amplitude, and is different from the thermal IR amplitude. This possibility was suggested by \citet{ahearn89} based on work by \citet{brown85} who showed that the amplitude of the thermal IR lightcurve may be dependent on both wavelength and elongation.


The varying lightcurve amplitudes will be utilized in Paper 2 when we determine the pole solution using the coma morphology from 1999 and 2010. As the viewing geometry varies during an apparition, the amplitude of the lightcurve will vary because different cross sections of the nucleus are observed. Thus, the lightcurve amplitudes will be useful in refining the pole solution and will help in constraining the shape of the nucleus.

We looked for evidence of periodicity in the coma signal by plotting both the total flux and flux per square arcsec in each photometric annulus as a function of rotational phase. What structure is visible on individual nights does not repeat in a coherent manner from night to night or from run to run. There was also no evidence of propagation of any features (either a maximum or minimum signal) outwards during a night. This is consistent with the analyses of \citet{jewitt89} and \citet{ahearn89} who did not see a rotation signal in their data.

\subsection{Lightcurve Asymmetry}
\label{lc_shape}
The rotational lightcurve in 1999 is somewhat asymmetric, allowing identification of certain features many rotations apart and eliminating $N - 0.5$, $N + 0.5$, $N + 1.5$... (where $N$ is the number of cycles) as possible solutions. This conclusively rules out the single- and triple-peaked solutions and reduces the uncertainty in the period determination since it halves the number of possible rotations. Any clear description of the lightcurve shape is made challenging because it is never observed completely on a given night. However, we note the following apparently repeated features: The minimum near 0.6 phase is deeper and steeper than the minimum near 0.1 phase when comparing lightcurves from the same observing run. There is a shoulder in the rising portion between 0.15 and 0.30 phase. Finally, while there are no nights in which both maxima were clearly observed, the maximum near 0.85 phase may be slightly higher than the maximum near 0.35 phase. We see evidence for each of these features in the 1988 data when discrete observing runs are considered\footnote{The 1988 April data from \citet{jewitt89} showed the deeper minimum to be at $\sim$0.7 phase, in contradiction to all other 1988 datasets (see Figure~\ref{fig:1988_data}). This dataset was the most difficult to normalize since neither April 9 or 12 contained a clear extremum, and the April 10 and 15 data required very different $\Delta$m$_2$ adjustments to bring the maxima in alignment near zero phase. Given that the minima were observed only five nights apart and no lightcurve change in shape is expected during this short a time interval, we are skeptical of this anomalous minimum.}. The features are also seen in the 1994 data from \citet{mueller96}, although since the viewing geometry was very different, we cannot be sure that they correspond to the same topographic features on the nucleus.

\section{Summary and Discussion}
\label{discussion}
We imaged Tempel 2 on 32 nights from 1999 April until 2000 March. We present here R-band nucleus lightcurves obtained on 11 of these nights from April through June (prior to perihelion), a total of 892 data points. Absolute calibrations and nightly extinction corrections were determined on photometric nights, and were applied to non-photometric nights to extract usable lightcurves. Field stars in the image were monitored on all nights and were used as comparison stars to correct the comet magnitude on non-photometric nights. A median coma was determined each night and removed from all images on the night. All magnitudes were normalized to $r$ = $\Delta$ = 1 AU and $\alpha$ = 0$^\circ$. A final adjustment, which accounts for uncertainties in the calibrations that tend to create nightly magnitude offsets such as the application of absolute calibrations on non-photometric nights, was applied to make each night peak at the same magnitude.

We determined a synodic rotation period of 8.941 $\pm$ 0.002 hr; no other double-peaked solutions exist. Our dataset is sufficient to rule out all aliases other than the single- and triple-peaked solutions, and these are ruled out by the differences in shape and brightness between the two minima. Our period matches one of the possible periods obtained by \citet{mueller96} from post-perihelion data in 1994 (during the same inter-perihelion time as our 1999 data) and rules out their other four possible solutions. We reanalyzed published data from 1988 using the same analysis techniques and found a synodic rotation period of 8.932 $\pm$ 0.001 hr. Our 1999 data cannot be fit by the rotation period from 1988, nor can the 1988 data be fit by the rotation period from 1999, even though the comet had nearly identical viewing geometries during the two sets of observations. Thus, we conclude that Tempel 2 spun-down by 0.009 hr ($\sim$32 s) from 1988 to 1999, an interval that included two perihelion passages.

This marks the first conclusive measurement of spin-down in a comet, although three comets have shown possible evidence of spin-up: 9P/Tempel 1 \citep{belton07,chesley10}, 2P/Encke \citep{fernandez05}, and Comet Levy (1990c) \citep{schleicher91,feldman92a}. Modeling by \citet{samarasinha95}, \citet{neishtadt02}, \citet{gutierrez03}, and others has shown that either spin-up or spin-down is possible, with spin-up more likely in the long run. These simulations have shown that under certain conditions the spin period can change by much larger amounts in a single orbit (e.g. typical changes of 0.01--10 hr per orbit according to \citet{gutierrez03}) than the measured change of 0.009 hr in two orbits for Tempel 2. Thus, the measured spin-down of 10P/Tempel 2 is probably not exceptional, and comparable changes in period could likely be measured in other comets if similarly high quality data were obtained over several epochs.

Possible causes of a change in spin-state are summarized by \citet{samarasinha04}.
The spin-down of Tempel 2 may have been caused by either a one-time event or by recurring torquing each orbit. Scenarios for causing a one-time change in rotation period include a short-lived, freshly exposed active region (causing temporary torquing), fragmentation (changing the moment of inertia), collision with another object, and tidal torquing from Jupiter. The orbit of Tempel 2 is well known and tidal torquing from Jupiter should not have been significant or more severe between 1988 and 1999 than on other recent orbits. Collisions are rare in the current solar system, and any collision significant enough to alter the rotation period of Tempel 2 would likely have resulted in an increase in brightness upon subsequent orbits, which has not been seen. Fragmentation is known to be relatively common, occurring on average at least once per century per comet (c.f. \citet{chen94,weissman80}). However, there is no evidence for Tempel 2 having split, such as a companion nucleus or a sustained increase in brightness as was observed in 73P/Schwassmann-Wachmann 3 \citep{ferrin10}. 
A new active region could be created by fragmention, impact, or outburst. Since outbursts are known to occur far more frequently than fragmentation or impacts, an outburst is the most plausible explanation for a one-time change in the rotation period. If the change in the period was caused by a unique event such as an outburst, it could have occurred anywhere in the orbit and we would expect no change in the rotation period on subsequent orbits.

It is challenging to devise a mechanism that is strong enough to affect the rotation period but lasts for a short time (less than one orbit). Thus, we suspect the spin-down is due to torques on the nucleus caused by asymmetric outgassing. This is believed to be the most common means of changing the spin state of a comet (c.f. \citet{samarasinha04}). Seasonal illumination of an active region could cause torquing which slows the rotation of the nucleus by a comparable amount each orbit. In the case of Tempel 2, the torquing likely occurs near or shortly after perihelion, when the comet is most active, although the length of time over which the torquing may occur is unknown. Since Tempel 2 reached perihelion twice between the 1988 and 1999 observations, in this, our preferred scenario, the rotation period must be changing by $\sim$16 s each orbit.


Tempel 2 reached perihelion on 2010 July 4, and we encourage its observation throughout the apparition to determine the rotation period. Due to a perturbation by Jupiter between the 1999 and 2005 apparitions, the viewing geometry is different in 2010 than in 1988 or 1999. The post-perihelion viewing geometry will be better this apparition and the comet will be visible longer each night. The smaller post-perihelion $\Delta$ in 2010 will provide better spatial resolution, making it easier to separate the nucleus and coma signals even though the coma is more extensive after perihelion. This may allow measurement of a nucleus lightcurve after perihelion, despite increased coma contamination relative to the pre-perihelion measurements in 1988, 1994, and 1999.  

Assuming the spin-down is recurrent each orbit, we predict that the pre-perihelion rotation period should be $\sim$32 s longer than in 1999 as there have been two intervening perihelion passages (1999 September 8 and 2005 February 15) since our data were obtained in 1999. If the pole position given by \citet{sekanina87b} is correct, the sidereal rotation period should be 8.950 $\pm$ 0.003 hr. Since it is unknown if the spin-down is caused by an instantaneous impulsive event or a steady change over an extended period of time, we cannot estimate the rotation period that might be observed shortly after perihelion in 2010. Observations long enough after perihelion to allow the torquing to have taken effect should reveal an additional lengthening of the sidereal rotation period by $\sim$16 s relative to the pre-perihelion 2010 period. Using again the pole solution of \citet{sekanina87b}, this would yield a sidereal period 8.954 $\pm$ 0.004 hr.

We have begun to observe Tempel 2 in 2010 and will combine these new data with our observations obtained during its active phase in 1999 as a follow up to the current paper. As of August 2010 (when the revised manuscript was submitted), we are unaware of any period determinations from 2010 observations. In Paper 2 we will use the different viewing geometry provided by the 2010 apparition to investigate the pole orientation and the location and activity of any jets. We hope to determine the rotation period in 2010, and possibly obtain additional post-perihelion observations after activity has subsided in 2011 or later to resolve whether the comet spins down each orbit. With its relatively low inclination ($i$ $\sim$ 12$^\circ$) and perihelion close to the Earth's orbit ($q$ $\sim$ 1.4 AU), Tempel 2 is frequently on the short list of targets for comet missions, making further study highly desirable.

\section*{Acknowledgements}
Thanks to Beatrice Mueller for a thorough and helpful review. We thank Christopher Henry, Kevin Walsh, and Wendy Williams for helping to obtain these observations, and Brian Skiff for useful discussions. We are grateful for JPL's Horizons for generating observing geometries. The period analysis was greatly facilitated by the use of the Data Desk exploratory data analysis package from Data Descriptions, Inc. MMK and DGS were supported by NASA Planetary Astronomy grant NNX09B51G. TLF was supported by multiple NASA grants. EWS was supported by an NSF grant to Northern Arizona University for the Research Experiences for Undergraduates program.



\clearpage

\renewcommand{\arraystretch}{0.6}

\begin{deluxetable}{lccccccccccccc}  
\tabletypesize{\scriptsize}
\tablecolumns{14}
\tablewidth{0pt} 
\setlength{\tabcolsep}{0.05in}
\tablecaption{Summary of Tempel 2 observations and geometric parameters in 1999.\,\tablenotemark{a}}
\tablehead{   
  \\[-1.0ex]
  \colhead{UT}&
  \colhead{UT}&
  \colhead{Tel.}&
  \colhead{CCD}&
  \colhead{$r$}&
  \colhead{$\Delta$}&
  \colhead{$\alpha$}&
  \colhead{$\Delta$t}&
  \colhead{Ecl. Long.}&
  \colhead{Ecl. Long.}&
  \colhead{$\Delta$m$_1$\,\tablenotemark{e}}&
  \colhead{$\Delta$m$_2$\,\tablenotemark{f}}&
  \colhead{$\sigma$$_{m_R}$}&
  \colhead{Conditions}\\
  \colhead{Date}&
   \colhead{Range}&
  \colhead{Diam.}&
  \colhead{}&
  \colhead{(AU)}&
  \colhead{(AU)}&
  \colhead{($^\circ$)}&
  \colhead{(hr)\,\tablenotemark{b}}&
  \colhead{Earth ($^\circ$)\,\tablenotemark{c}}&
  \colhead{Sun ($^\circ$)\,\tablenotemark{d}}&
  \colhead{}&
  \colhead{}&
  \colhead{}&
  \colhead{}
}
\startdata
\\[-1.0ex]
Apr 17&7:48--11:55&1.8-m&SITe&2.025&1.275&24.0&0.177&80.1&56.3&$-$2.84&$-$0.02&0.02--0.04&thin clouds\\
Apr 18&7:20--12:11&1.8-m&SITe&2.020&1.262&23.8&0.176&80.3&56.7&$-$2.81&+0.08&0.02--0.06&thin clouds\\
Apr 19&6:53--12:10&1.8-m&SITe&2.014&1.248&23.7&0.174&80.5&57.1&$-$2.77&$+$0.03&0.03&photometric\\
May 26&6:37--11:22&1.8-m&SITe&1.807&0.839&13.9&0.117&81.9&72.2&$-$1.36&+0.15&0.01&thin clouds\\
May 27&7:43--11:30&1.8-m&SITe&1.802&0.831&13.6&0.116&81.8&72.7&$-$1.33&+0.26&0.01&thin clouds\\
Jun 08&7:05--11:14&1.1-m&TI&1.741&0.748&10.7&0.104&80.1&78.4&$-$0.92&...&0.02&photometric\\
Jun 09&4:45--\phantom{0}9:36&1.1-m&TI&1.736&0.743&10.6&0.103&80.0&78.8&$-$0.90&...&0.03&photometric\\
Jun 10&4:34--11:17&1.1-m&TI&1.731&0.737&10.5&0.102&79.7&79.3&$-$0.87&$-$0.02&0.02&photometric\\
Jun 11&4:30--\phantom{0}9:35&1.1-m&TI&1.726&0.732&10.4&0.102&79.6&79.8&$-$0.85&...&0.02&photometric\\
Jun 22&7:29--11:08&1.1-m&TI&1.675&0.685&12.4&0.095&77.5&85.4&$-$0.69&$-$0.07&0.02--0.06&thin clouds\\
Jun 23&3:57--11:04&1.1-m&TI&1.671&0.682&12.7&0.095&77.4&85.9&$-$0.69&$-$0.07&0.02--0.04&photometric\\
\enddata
\tablenotetext{a} {All parameters were taken at the midpoint of each night's observations, and all images were obtained at Lowell Observatory.}
\tablenotetext{b}{Light travel time.}
\tablenotetext{c}{Ecliptic longitude of the Earth as seen from the comet.}
\tablenotetext{d}{Ecliptic longitude of the Sun as seen from the comet.}
\tablenotetext{e}{Magnitude necessary to correct from m$_R$($r$,$\Delta$,$\alpha$) to m$_R$(1,1,0) for the night (in magnitudes).}
\tablenotetext{f}{Offset (in magnitudes) necessary to make data on all nights peak at the same magnitude.}
\label{t:obs_summary}
\end{deluxetable}

\clearpage



\renewcommand{\thefootnote}{\alph{footnote}}
\scriptsize
\begin{center}
\begin{longtable}{lccc}
\caption[Table of photometry]{Table of photometry}\label{t:photometry}\\

\hline
\hline\\[-1.0ex]
  \multicolumn{1}{l}{UT Date}&
  \multicolumn{1}{c}{UT\,\tablenotemark{a}}&
  \multicolumn{1}{c}{m$_R$\,\tablenotemark{b}}&
  \multicolumn{1}{c}{m$_R$*\,\tablenotemark{c}}\\[0.8ex]
  \hline\\[-1.0ex]

\endfirsthead

\multicolumn{4}{c}%
{{\bfseries \tablename\ \thetable{} -- continued from previous page}} \\[0.8ex]
\hline\\[-1.0ex]
\endhead


\endlastfoot

Apr 17&\phantom{0}7.848&16.93&14.38\\ 
Apr 17&\phantom{0}7.915&16.90&14.34\\ 
Apr 17&\phantom{0}7.952&17.00&14.47\\ 
Apr 17&\phantom{0}7.989&16.90&14.34\\ 
Apr 17&\phantom{0}8.038&16.83&14.25\\ 
Apr 17&\phantom{0}8.076&16.89&14.33\\ 
Apr 17&\phantom{0}8.112&16.84&14.26\\ 
Apr 17&\phantom{0}8.201&16.76&14.16\\ 
Apr 17&\phantom{0}8.294&16.83&14.24\\ 
Apr 17&\phantom{0}8.331&16.74&14.13\\ 
Apr 17&\phantom{0}8.368&16.73&14.13\\ 
Apr 17&\phantom{0}8.421&16.73&14.13\\ 
Apr 17&\phantom{0}8.458&16.68&14.06\\ 
Apr 17&\phantom{0}8.495&16.69&14.07\\ 
Apr 17&\phantom{0}8.547&16.72&14.11\\ 
Apr 17&\phantom{0}8.637&16.71&14.10\\ 
Apr 17&\phantom{0}8.674&16.68&14.07\\ 
Apr 17&\phantom{0}8.711&16.73&14.12\\ 
Apr 17&\phantom{0}8.748&16.71&14.10\\ 
Apr 17&\phantom{0}8.785&16.67&14.04\\ 
Apr 17&\phantom{0}8.821&16.63&14.00\\ 
Apr 17&\phantom{0}8.893&16.64&14.02\\ 
Apr 17&\phantom{0}8.972&16.65&14.02\\ 
Apr 17&\phantom{0}9.067&16.62&13.99\\ 
Apr 17&\phantom{0}9.104&16.59&13.94\\ 
Apr 17&\phantom{0}9.142&16.60&13.96\\ 
Apr 17&\phantom{0}9.179&16.58&13.93\\ 
Apr 17&\phantom{0}9.216&16.61&13.98\\ 
Apr 17&\phantom{0}9.253&16.58&13.93\\ 
Apr 17&\phantom{0}9.315&16.58&13.93\\ 
Apr 17&\phantom{0}9.421&16.56&13.91\\ 
Apr 17&\phantom{0}9.486&16.55&13.91\\ 
Apr 17&\phantom{0}9.523&16.54&13.90\\ 
Apr 17&\phantom{0}9.560&16.56&13.92\\ 
Apr 17&\phantom{0}9.597&16.56&13.91\\ 
Apr 17&\phantom{0}9.634&16.59&13.96\\ 
Apr 17&\phantom{0}9.671&16.57&13.93\\ 
Apr 17&\phantom{0}9.716&16.60&13.96\\ 
Apr 17&\phantom{0}9.807&16.60&13.97\\ 
Apr 17&\phantom{0}9.844&16.63&14.00\\ 
Apr 17&\phantom{0}9.881&16.62&13.98\\ 
Apr 17&\phantom{0}9.918&16.62&13.99\\ 
Apr 17&\phantom{0}9.955&16.62&13.99\\ 
Apr 17&\phantom{0}9.992&16.62&13.99\\ 
Apr 17&10.037&16.64&14.01\\ 
Apr 17&10.219&16.67&14.05\\ 
Apr 17&10.288&16.70&14.09\\ 
Apr 17&10.325&16.68&14.07\\ 
Apr 17&10.362&16.71&14.11\\ 
Apr 17&10.399&16.73&14.13\\ 
Apr 17&10.436&16.75&14.14\\ 
Apr 17&10.473&16.75&14.16\\ 
Apr 17&10.604&16.79&14.21\\ 
Apr 17&11.689&16.95&14.41\\ 
Apr 17&11.726&16.95&14.41\\ 
Apr 17&11.763&16.93&14.38\\ 
Apr 17&11.800&16.94&14.39\\ 
Apr 17&11.890&16.91&14.35\\ 
\\
Apr 18&\phantom{0}7.681&16.47&13.89\\ 
Apr 18&\phantom{0}7.718&16.44&13.86\\ 
Apr 18&\phantom{0}7.755&16.46&13.89\\ 
Apr 18&\phantom{0}7.792&16.48&13.91\\ 
Apr 18&\phantom{0}7.829&16.48&13.90\\ 
Apr 18&\phantom{0}7.867&16.47&13.90\\ 
Apr 18&\phantom{0}7.906&16.49&13.91\\ 
Apr 18&\phantom{0}7.943&16.51&13.94\\ 
Apr 18&\phantom{0}7.980&16.55&13.99\\ 
Apr 18&\phantom{0}8.017&16.55&13.99\\ 
Apr 18&\phantom{0}8.054&16.61&14.05\\ 
Apr 18&\phantom{0}8.091&16.62&14.06\\ 
Apr 18&\phantom{0}8.160&16.61&14.05\\ 
Apr 18&\phantom{0}8.234&16.63&14.08\\ 
Apr 18&\phantom{0}8.395&16.68&14.14\\ 
Apr 18&\phantom{0}8.432&16.69&14.15\\ 
Apr 18&\phantom{0}8.469&16.74&14.21\\ 
Apr 18&\phantom{0}8.506&16.66&14.12\\ 
Apr 18&\phantom{0}8.543&16.78&14.25\\ 
Apr 18&\phantom{0}8.580&16.73&14.20\\ 
Apr 18&\phantom{0}8.621&16.74&14.21\\ 
Apr 18&\phantom{0}8.702&16.83&14.31\\ 
Apr 18&\phantom{0}8.739&16.81&14.29\\ 
Apr 18&\phantom{0}8.777&16.84&14.33\\ 
Apr 18&\phantom{0}8.814&16.83&14.32\\ 
Apr 18&\phantom{0}8.851&16.89&14.39\\ 
Apr 18&\phantom{0}8.888&16.90&14.40\\ 
Apr 18&\phantom{0}8.962&16.93&14.44\\ 
Apr 18&\phantom{0}9.028&16.98&14.50\\ 
Apr 18&\phantom{0}9.109&17.00&14.53\\ 
Apr 18&\phantom{0}9.146&16.99&14.52\\ 
Apr 18&\phantom{0}9.183&17.04&14.57\\ 
Apr 18&\phantom{0}9.220&17.01&14.54\\ 
Apr 18&\phantom{0}9.258&17.03&14.56\\ 
Apr 18&\phantom{0}9.295&17.05&14.59\\ 
Apr 18&\phantom{0}9.332&17.07&14.61\\ 
Apr 18&\phantom{0}9.429&17.07&14.61\\ 
Apr 18&\phantom{0}9.535&17.06&14.60\\ 
Apr 18&\phantom{0}9.611&17.02&14.55\\ 
Apr 18&\phantom{0}9.649&17.03&14.56\\ 
Apr 18&\phantom{0}9.686&17.01&14.54\\ 
Apr 18&\phantom{0}9.723&17.01&14.53\\ 
Apr 18&\phantom{0}9.760&17.04&14.57\\ 
Apr 18&\phantom{0}9.797&17.07&14.61\\ 
Apr 18&\phantom{0}9.836&17.04&14.58\\ 
Apr 18&\phantom{0}9.873&17.01&14.54\\ 
Apr 18&\phantom{0}9.910&17.01&14.54\\ 
Apr 18&\phantom{0}9.947&16.98&14.51\\ 
Apr 18&\phantom{0}9.985&16.98&14.50\\ 
Apr 18&10.022&16.98&14.50\\ 
Apr 18&10.062&16.99&14.51\\ 
Apr 18&10.391&16.73&14.20\\ 
Apr 18&10.460&16.86&14.35\\ 
Apr 18&10.541&16.87&14.37\\ 
Apr 18&10.579&16.85&14.35\\ 
Apr 18&10.616&16.83&14.32\\ 
Apr 18&10.653&16.80&14.28\\ 
Apr 18&10.690&16.90&14.40\\ 
Apr 18&10.811&16.80&14.29\\ 
Apr 18&10.917&16.75&14.22\\ 
Apr 18&10.959&16.77&14.24\\ 
Apr 18&10.996&16.67&14.13\\ 
Apr 18&11.033&16.74&14.22\\ 
Apr 18&11.070&16.74&14.21\\ 
Apr 18&11.107&16.71&14.18\\ 
Apr 18&11.144&16.70&14.16\\ 
Apr 18&11.210&16.68&14.14\\ 
Apr 18&11.277&16.67&14.13\\ 
Apr 18&11.356&16.62&14.07\\ 
Apr 18&11.393&16.64&14.09\\ 
Apr 18&11.431&16.64&14.09\\ 
Apr 18&11.468&16.66&14.12\\ 
Apr 18&11.505&16.61&14.06\\ 
Apr 18&11.542&16.61&14.06\\ 
Apr 18&11.623&16.60&14.05\\ 
Apr 18&11.664&16.57&14.01\\ 
Apr 18&11.702&16.57&14.01\\ 
Apr 18&11.739&16.58&14.02\\ 
Apr 18&11.776&16.56&14.01\\ 
Apr 18&11.813&16.55&13.99\\ 
Apr 18&11.851&16.54&13.98\\ 
Apr 18&11.902&16.53&13.96\\ 
\\
Apr 19&\phantom{0}7.129&16.77&14.22\\ 
Apr 19&\phantom{0}7.170&16.80&14.25\\ 
Apr 19&\phantom{0}7.208&16.74&14.18\\ 
Apr 19&\phantom{0}7.824&17.07&14.57\\ 
Apr 19&\phantom{0}7.889&17.06&14.57\\ 
Apr 19&\phantom{0}7.956&17.08&14.59\\ 
Apr 19&\phantom{0}7.993&17.06&14.56\\ 
Apr 19&\phantom{0}8.031&17.09&14.61\\ 
Apr 19&\phantom{0}8.068&17.12&14.64\\ 
Apr 19&\phantom{0}8.105&17.07&14.58\\ 
Apr 19&\phantom{0}8.142&17.10&14.62\\ 
Apr 19&\phantom{0}8.185&17.07&14.59\\ 
Apr 19&\phantom{0}8.281&17.10&14.62\\ 
Apr 19&\phantom{0}8.319&17.02&14.52\\ 
Apr 19&\phantom{0}8.356&16.99&14.48\\ 
Apr 19&\phantom{0}8.393&17.02&14.52\\ 
Apr 19&\phantom{0}8.430&16.96&14.45\\ 
Apr 19&\phantom{0}8.467&17.03&14.53\\ 
Apr 19&\phantom{0}8.606&16.93&14.41\\ 
Apr 19&\phantom{0}8.692&16.82&14.27\\ 
Apr 19&\phantom{0}8.730&16.87&14.34\\ 
Apr 19&\phantom{0}8.766&16.82&14.28\\ 
Apr 19&\phantom{0}8.804&16.86&14.33\\ 
Apr 19&\phantom{0}8.841&16.83&14.29\\ 
Apr 19&\phantom{0}8.878&16.87&14.34\\ 
Apr 19&\phantom{0}9.987&16.49&13.89\\ 
Apr 19&10.024&16.50&13.91\\ 
Apr 19&10.062&16.48&13.88\\ 
Apr 19&10.099&16.50&13.90\\ 
Apr 19&10.136&16.49&13.90\\ 
Apr 19&10.172&16.49&13.90\\ 
Apr 19&10.211&16.49&13.90\\ 
Apr 19&10.292&16.51&13.91\\ 
Apr 19&10.329&16.48&13.89\\ 
Apr 19&10.366&16.48&13.88\\ 
Apr 19&10.384&16.48&13.88\\ 
Apr 19&10.441&16.48&13.88\\ 
Apr 19&10.478&16.49&13.89\\ 
Apr 19&10.578&16.48&13.88\\ 
Apr 19&10.658&16.49&13.90\\ 
Apr 19&10.696&16.50&13.91\\ 
Apr 19&10.732&16.51&13.92\\ 
Apr 19&10.769&16.51&13.92\\ 
Apr 19&10.806&16.50&13.91\\ 
Apr 19&10.844&16.51&13.92\\ 
Apr 19&10.907&16.53&13.94\\ 
Apr 19&10.974&16.53&13.95\\ 
Apr 19&11.889&16.89&14.36\\ 
Apr 19&11.926&16.87&14.35\\ 
Apr 19&11.964&16.91&14.39\\ 
Apr 19&12.001&17.01&14.54\\ 
Apr 19&12.038&16.92&14.44\\ 
Apr 19&12.076&16.92&14.44\\ 
Apr 19&12.113&16.95&14.47\\ 
Apr 19&12.151&17.01&14.55\\ 
\\
May 26&\phantom{0}6.640&14.91&13.99\\ 
May 26&\phantom{0}6.679&14.89&13.97\\ 
May 26&\phantom{0}6.716&14.88&13.95\\ 
May 26&\phantom{0}6.760&14.87&13.95\\ 
May 26&\phantom{0}6.801&14.88&13.95\\ 
May 26&\phantom{0}6.838&14.86&13.93\\ 
May 26&\phantom{0}6.875&14.85&13.92\\ 
May 26&\phantom{0}6.912&14.85&13.92\\ 
May 26&\phantom{0}6.992&14.84&13.91\\ 
May 26&\phantom{0}7.029&14.84&13.91\\ 
May 26&\phantom{0}7.066&14.84&13.90\\ 
May 26&\phantom{0}7.103&14.82&13.89\\ 
May 26&\phantom{0}7.140&14.83&13.90\\ 
May 26&\phantom{0}7.191&14.84&13.91\\ 
May 26&\phantom{0}7.227&14.83&13.90\\ 
May 26&\phantom{0}7.264&14.84&13.90\\ 
May 26&\phantom{0}7.302&14.83&13.89\\ 
May 26&\phantom{0}7.339&14.81&13.87\\ 
May 26&\phantom{0}7.517&14.82&13.88\\ 
May 26&\phantom{0}7.554&14.82&13.88\\ 
May 26&\phantom{0}7.591&14.81&13.88\\ 
May 26&\phantom{0}7.631&14.82&13.88\\ 
May 26&\phantom{0}7.668&14.82&13.88\\ 
May 26&\phantom{0}7.705&14.82&13.89\\ 
May 26&\phantom{0}7.742&14.82&13.88\\ 
May 26&\phantom{0}7.779&14.84&13.90\\ 
May 26&\phantom{0}7.819&14.83&13.89\\ 
May 26&\phantom{0}7.857&14.84&13.91\\ 
May 26&\phantom{0}7.894&14.85&13.92\\ 
May 26&\phantom{0}7.931&14.85&13.91\\ 
May 26&\phantom{0}7.968&14.85&13.93\\ 
May 26&\phantom{0}8.935&15.10&14.25\\ 
May 26&\phantom{0}8.974&15.13&14.30\\ 
May 26&\phantom{0}9.011&15.12&14.28\\ 
May 26&\phantom{0}9.048&15.12&14.29\\ 
May 26&\phantom{0}9.085&15.14&14.31\\ 
May 26&\phantom{0}9.122&15.13&14.30\\ 
May 26&\phantom{0}9.163&15.13&14.30\\ 
May 26&\phantom{0}9.200&15.14&14.31\\ 
May 26&\phantom{0}9.237&15.14&14.31\\ 
May 26&\phantom{0}9.274&15.13&14.30\\ 
May 26&10.044&15.12&14.29\\ 
May 26&10.081&15.11&14.27\\ 
May 26&10.119&15.11&14.26\\ 
May 26&10.156&15.10&14.26\\ 
May 26&10.195&15.08&14.23\\ 
May 26&10.232&15.07&14.21\\ 
May 26&10.269&15.05&14.18\\ 
May 26&10.307&15.04&14.18\\ 
May 26&10.344&15.03&14.17\\ 
May 26&10.484&15.01&14.13\\ 
May 26&10.521&14.99&14.11\\ 
May 26&10.558&15.00&14.11\\ 
May 26&10.596&15.00&14.11\\ 
May 26&10.632&14.98&14.10\\ 
May 26&11.187&14.94&14.04\\ 
May 26&11.225&14.94&14.04\\ 
May 26&11.262&14.92&14.01\\ 
May 26&11.299&14.92&14.02\\ 
May 26&11.353&14.92&14.02\\ 
\\
May 27&\phantom{0}7.732&15.12&14.33\\ 
May 27&\phantom{0}7.777&15.12&14.33\\ 
May 27&\phantom{0}7.820&15.13&14.34\\ 
May 27&\phantom{0}7.859&15.15&14.36\\ 
May 27&\phantom{0}7.899&15.14&14.36\\ 
May 27&\phantom{0}7.936&15.14&14.35\\ 
May 27&\phantom{0}7.974&15.15&14.36\\ 
May 27&\phantom{0}8.012&15.15&14.36\\ 
May 27&\phantom{0}8.050&15.14&14.36\\ 
May 27&\phantom{0}8.088&15.15&14.36\\ 
May 27&\phantom{0}8.129&15.15&14.37\\ 
May 27&\phantom{0}8.168&15.14&14.35\\ 
May 27&\phantom{0}8.206&15.13&14.34\\ 
May 27&\phantom{0}8.243&15.13&14.34\\ 
May 27&\phantom{0}8.280&15.13&14.34\\ 
May 27&\phantom{0}8.317&15.12&14.32\\ 
May 27&\phantom{0}8.356&15.12&14.32\\ 
May 27&\phantom{0}8.394&15.10&14.30\\ 
May 27&\phantom{0}8.431&15.10&14.30\\ 
May 27&\phantom{0}8.468&15.10&14.30\\ 
May 27&\phantom{0}8.505&15.08&14.28\\ 
May 27&\phantom{0}8.544&15.07&14.27\\ 
May 27&\phantom{0}8.581&15.06&14.25\\ 
May 27&\phantom{0}8.618&15.04&14.23\\ 
May 27&\phantom{0}8.655&15.03&14.22\\ 
May 27&\phantom{0}8.692&15.03&14.21\\ 
May 27&\phantom{0}8.734&15.02&14.20\\ 
May 27&\phantom{0}8.771&15.01&14.19\\ 
May 27&\phantom{0}8.808&15.01&14.18\\ 
May 27&\phantom{0}8.846&15.00&14.18\\ 
May 27&\phantom{0}8.883&15.00&14.18\\ 
May 27&\phantom{0}9.303&14.87&14.02\\ 
May 27&\phantom{0}9.340&14.88&14.03\\ 
May 27&\phantom{0}9.377&14.87&14.02\\ 
May 27&\phantom{0}9.414&14.87&14.02\\ 
May 27&\phantom{0}9.451&14.86&14.00\\ 
May 27&\phantom{0}9.915&14.78&13.91\\ 
May 27&\phantom{0}9.952&14.78&13.91\\ 
May 27&\phantom{0}9.989&14.77&13.90\\ 
May 27&10.026&14.77&13.90\\ 
May 27&10.063&14.77&13.91\\ 
May 27&10.137&14.75&13.88\\ 
May 27&10.175&14.75&13.88\\ 
May 27&10.212&14.75&13.88\\ 
May 27&10.249&14.74&13.87\\ 
May 27&10.291&14.75&13.88\\ 
May 27&10.366&14.75&13.88\\ 
May 27&10.402&14.75&13.88\\ 
May 27&10.489&14.76&13.89\\ 
May 27&10.526&14.77&13.90\\ 
May 27&10.563&14.76&13.90\\ 
May 27&11.425&14.91&14.07\\ 
\\
Jun 08&\phantom{0}7.113&14.61&14.08\\ 
Jun 08&\phantom{0}7.148&14.60&14.07\\ 
Jun 08&\phantom{0}7.184&14.59&14.06\\ 
Jun 08&\phantom{0}7.219&14.59&14.05\\ 
Jun 08&\phantom{0}7.254&14.58&14.04\\ 
Jun 08&\phantom{0}7.293&14.56&14.01\\ 
Jun 08&\phantom{0}7.328&14.55&13.99\\ 
Jun 08&\phantom{0}7.363&14.53&13.98\\ 
Jun 08&\phantom{0}7.398&14.52&13.96\\ 
Jun 08&\phantom{0}7.434&14.53&13.96\\ 
Jun 08&\phantom{0}7.568&14.50&13.93\\ 
Jun 08&\phantom{0}7.643&14.49&13.92\\ 
Jun 08&\phantom{0}7.680&14.49&13.92\\ 
Jun 08&\phantom{0}7.814&14.48&13.90\\ 
Jun 08&\phantom{0}7.851&14.48&13.90\\ 
Jun 08&\phantom{0}7.890&14.49&13.92\\ 
Jun 08&\phantom{0}8.024&14.49&13.91\\ 
Jun 08&\phantom{0}8.060&14.49&13.91\\ 
Jun 08&\phantom{0}8.095&14.49&13.92\\ 
Jun 08&\phantom{0}8.134&14.48&13.89\\ 
Jun 08&\phantom{0}8.169&14.47&13.89\\ 
Jun 08&\phantom{0}8.204&14.47&13.88\\ 
Jun 08&\phantom{0}8.239&14.47&13.89\\ 
Jun 08&\phantom{0}8.275&14.47&13.88\\ 
Jun 08&\phantom{0}8.312&14.47&13.89\\ 
Jun 08&\phantom{0}8.347&14.46&13.87\\ 
Jun 08&\phantom{0}8.382&14.47&13.88\\ 
Jun 08&\phantom{0}8.418&14.47&13.89\\ 
Jun 08&\phantom{0}8.453&14.47&13.89\\ 
Jun 08&\phantom{0}8.490&14.46&13.88\\ 
Jun 08&\phantom{0}8.525&14.46&13.87\\ 
Jun 08&\phantom{0}8.560&14.47&13.88\\ 
Jun 08&\phantom{0}8.596&14.47&13.89\\ 
Jun 08&\phantom{0}8.631&14.48&13.90\\ 
Jun 08&\phantom{0}8.676&14.48&13.91\\ 
Jun 08&\phantom{0}8.711&14.49&13.91\\ 
Jun 08&\phantom{0}8.746&14.48&13.90\\ 
Jun 08&\phantom{0}8.781&14.49&13.91\\ 
Jun 08&\phantom{0}8.816&14.50&13.93\\ 
Jun 08&\phantom{0}9.065&14.57&14.02\\ 
Jun 08&\phantom{0}9.100&14.56&14.01\\ 
Jun 08&\phantom{0}9.135&14.57&14.03\\ 
Jun 08&\phantom{0}9.170&14.57&14.02\\ 
Jun 08&\phantom{0}9.205&14.59&14.05\\ 
Jun 08&\phantom{0}9.489&14.66&14.15\\ 
Jun 08&\phantom{0}9.525&14.66&14.15\\ 
Jun 08&\phantom{0}9.560&14.67&14.16\\ 
Jun 08&\phantom{0}9.595&14.67&14.17\\ 
Jun 08&\phantom{0}9.630&14.70&14.21\\ 
Jun 08&\phantom{0}9.908&14.72&14.25\\ 
Jun 08&\phantom{0}9.943&14.72&14.24\\ 
Jun 08&\phantom{0}9.978&14.73&14.26\\ 
Jun 08&10.013&14.74&14.27\\ 
Jun 08&10.048&14.78&14.33\\ 
Jun 08&10.194&14.76&14.31\\ 
Jun 08&10.229&14.77&14.32\\ 
Jun 08&10.265&14.77&14.31\\ 
Jun 08&10.300&14.77&14.31\\ 
Jun 08&10.335&14.77&14.32\\ 
Jun 08&10.467&14.74&14.27\\ 
Jun 08&10.502&14.76&14.30\\ 
Jun 08&10.538&14.74&14.28\\ 
Jun 08&10.573&14.76&14.30\\ 
Jun 08&10.608&14.78&14.33\\ 
Jun 08&10.646&14.77&14.33\\ 
Jun 08&10.681&14.76&14.31\\ 
Jun 08&10.716&14.74&14.27\\ 
Jun 08&10.751&14.74&14.28\\ 
Jun 08&10.786&14.74&14.28\\ 
Jun 08&10.837&14.74&14.27\\ 
Jun 08&10.872&14.74&14.27\\ 
Jun 08&10.907&14.75&14.29\\ 
Jun 08&10.942&14.74&14.28\\ 
Jun 08&10.977&14.74&14.27\\ 
Jun 08&11.160&14.70&14.21\\ 
Jun 08&11.195&14.69&14.21\\ 
Jun 08&11.230&14.71&14.23\\ 
\\
Jun 09&\phantom{0}4.315&14.76&14.35\\ 
Jun 09&\phantom{0}4.342&14.74&14.32\\ 
Jun 09&\phantom{0}4.368&14.74&14.31\\ 
Jun 09&\phantom{0}4.395&14.70&14.25\\ 
Jun 09&\phantom{0}4.422&14.68&14.22\\ 
Jun 09&\phantom{0}4.451&14.69&14.23\\ 
Jun 09&\phantom{0}4.478&14.69&14.24\\ 
Jun 09&\phantom{0}4.505&14.69&14.23\\ 
Jun 09&\phantom{0}4.531&14.67&14.20\\ 
Jun 09&\phantom{0}4.558&14.72&14.29\\ 
Jun 09&\phantom{0}4.764&14.71&14.27\\ 
Jun 09&\phantom{0}4.790&14.72&14.28\\ 
Jun 09&\phantom{0}4.817&14.73&14.30\\ 
Jun 09&\phantom{0}4.844&14.71&14.26\\ 
Jun 09&\phantom{0}4.871&14.71&14.27\\ 
Jun 09&\phantom{0}4.900&14.71&14.27\\ 
Jun 09&\phantom{0}4.926&14.70&14.26\\ 
Jun 09&\phantom{0}4.953&14.70&14.25\\ 
Jun 09&\phantom{0}4.980&14.69&14.23\\ 
Jun 09&\phantom{0}5.007&14.69&14.24\\ 
Jun 09&\phantom{0}5.452&14.64&14.16\\ 
Jun 09&\phantom{0}5.479&14.63&14.15\\ 
Jun 09&\phantom{0}5.506&14.63&14.15\\ 
Jun 09&\phantom{0}5.533&14.62&14.14\\ 
Jun 09&\phantom{0}5.560&14.59&14.09\\ 
Jun 09&\phantom{0}5.587&14.59&14.09\\ 
Jun 09&\phantom{0}5.614&14.58&14.08\\ 
Jun 09&\phantom{0}5.641&14.58&14.07\\ 
Jun 09&\phantom{0}5.668&14.57&14.06\\ 
Jun 09&\phantom{0}5.694&14.57&14.06\\ 
Jun 09&\phantom{0}5.986&14.56&14.04\\ 
Jun 09&\phantom{0}6.012&14.55&14.03\\ 
Jun 09&\phantom{0}6.039&14.54&14.03\\ 
Jun 09&\phantom{0}6.066&14.54&14.02\\ 
Jun 09&\phantom{0}6.093&14.54&14.03\\ 
Jun 09&\phantom{0}6.120&14.55&14.03\\ 
Jun 09&\phantom{0}6.147&14.52&14.00\\ 
Jun 09&\phantom{0}6.174&14.52&14.00\\ 
Jun 09&\phantom{0}6.201&14.54&14.02\\ 
Jun 09&\phantom{0}6.457&14.50&13.97\\ 
Jun 09&\phantom{0}6.484&14.50&13.96\\ 
Jun 09&\phantom{0}6.511&14.49&13.96\\ 
Jun 09&\phantom{0}6.538&14.49&13.95\\ 
Jun 09&\phantom{0}6.565&14.49&13.95\\ 
Jun 09&\phantom{0}6.592&14.48&13.94\\ 
Jun 09&\phantom{0}7.256&14.47&13.92\\ 
Jun 09&\phantom{0}7.283&14.47&13.92\\ 
Jun 09&\phantom{0}7.310&14.48&13.94\\ 
Jun 09&\phantom{0}7.337&14.49&13.94\\ 
Jun 09&\phantom{0}7.364&14.49&13.95\\ 
Jun 09&\phantom{0}7.391&14.50&13.96\\ 
Jun 09&\phantom{0}7.418&14.49&13.96\\ 
Jun 09&\phantom{0}7.444&14.51&13.97\\ 
Jun 09&\phantom{0}7.471&14.49&13.96\\ 
Jun 09&\phantom{0}7.498&14.49&13.96\\ 
Jun 09&\phantom{0}8.011&14.62&14.14\\ 
Jun 09&\phantom{0}8.038&14.63&14.15\\ 
Jun 09&\phantom{0}8.065&14.63&14.15\\ 
Jun 09&\phantom{0}8.092&14.62&14.14\\ 
Jun 09&\phantom{0}8.119&14.63&14.15\\ 
Jun 09&\phantom{0}8.146&14.63&14.15\\ 
Jun 09&\phantom{0}8.173&14.64&14.16\\ 
Jun 09&\phantom{0}8.199&14.66&14.19\\ 
Jun 09&\phantom{0}8.226&14.65&14.17\\ 
Jun 09&\phantom{0}8.253&14.66&14.19\\ 
Jun 09&\phantom{0}8.494&14.70&14.25\\ 
Jun 09&\phantom{0}8.521&14.69&14.24\\ 
Jun 09&\phantom{0}8.548&14.72&14.28\\ 
Jun 09&\phantom{0}8.575&14.71&14.27\\ 
Jun 09&\phantom{0}8.602&14.71&14.27\\ 
Jun 09&\phantom{0}8.629&14.71&14.27\\ 
Jun 09&\phantom{0}8.656&14.72&14.28\\ 
Jun 09&\phantom{0}8.682&14.72&14.28\\ 
Jun 09&\phantom{0}8.709&14.72&14.29\\ 
Jun 09&\phantom{0}8.736&14.72&14.29\\ 
Jun 09&\phantom{0}8.768&14.74&14.31\\ 
Jun 09&\phantom{0}8.795&14.74&14.31\\ 
Jun 09&\phantom{0}8.822&14.72&14.29\\ 
Jun 09&\phantom{0}8.848&14.73&14.31\\ 
Jun 09&\phantom{0}8.875&14.74&14.31\\ 
Jun 09&\phantom{0}8.902&14.74&14.32\\ 
Jun 09&\phantom{0}8.929&14.73&14.30\\ 
Jun 09&\phantom{0}8.956&14.73&14.30\\ 
Jun 09&\phantom{0}8.983&14.73&14.30\\ 
Jun 09&\phantom{0}9.010&14.73&14.30\\ 
Jun 09&\phantom{0}9.346&14.68&14.22\\ 
Jun 09&\phantom{0}9.373&14.67&14.22\\ 
Jun 09&\phantom{0}9.400&14.71&14.27\\ 
Jun 09&\phantom{0}9.427&14.66&14.21\\ 
Jun 09&\phantom{0}9.454&14.66&14.20\\ 
Jun 09&\phantom{0}9.481&14.69&14.25\\ 
Jun 09&\phantom{0}9.507&14.69&14.25\\ 
Jun 09&\phantom{0}9.534&14.69&14.25\\ 
Jun 09&\phantom{0}9.561&14.68&14.22\\ 
Jun 09&\phantom{0}9.588&14.67&14.21\\ 
Jun 09&10.269&14.47&13.93\\ 
Jun 09&10.561&14.51&13.98\\ 
Jun 09&10.588&14.55&14.03\\ 
Jun 09&10.615&14.49&13.96\\ 
Jun 09&10.642&14.48&13.94\\ 
Jun 09&10.669&14.45&13.90\\ 
Jun 09&10.699&14.47&13.93\\ 
Jun 09&10.726&14.47&13.93\\ 
Jun 09&10.753&14.46&13.91\\ 
Jun 09&10.780&14.46&13.91\\ 
Jun 09&10.807&14.44&13.89\\ 
Jun 09&10.867&14.47&13.92\\ 
Jun 09&10.925&14.44&13.89\\ 
Jun 09&10.952&14.42&13.87\\ 
Jun 09&10.978&14.45&13.91\\ 
Jun 09&11.005&14.45&13.90\\ 
Jun 09&11.032&14.45&13.90\\ 
Jun 09&11.066&14.43&13.88\\ 
Jun 09&11.093&14.43&13.88\\ 
Jun 09&11.120&14.44&13.89\\ 
\\
Jun 10&\phantom{0}4.648&14.48&13.93\\ 
Jun 10&\phantom{0}4.708&14.48&13.93\\ 
Jun 10&\phantom{0}4.735&14.47&13.92\\ 
Jun 10&\phantom{0}4.762&14.47&13.92\\ 
Jun 10&\phantom{0}4.796&14.47&13.92\\ 
Jun 10&\phantom{0}4.829&14.44&13.88\\ 
Jun 10&\phantom{0}4.856&14.44&13.88\\ 
Jun 10&\phantom{0}4.882&14.45&13.90\\ 
Jun 10&\phantom{0}4.909&14.45&13.90\\ 
Jun 10&\phantom{0}4.936&14.46&13.91\\ 
Jun 10&\phantom{0}4.963&14.42&13.85\\ 
Jun 10&\phantom{0}4.990&14.44&13.88\\ 
Jun 10&\phantom{0}5.017&14.44&13.87\\ 
Jun 10&\phantom{0}5.044&14.44&13.87\\ 
Jun 10&\phantom{0}5.070&14.45&13.89\\ 
Jun 10&\phantom{0}5.249&14.49&13.95\\ 
Jun 10&\phantom{0}5.276&14.47&13.92\\ 
Jun 10&\phantom{0}5.303&14.45&13.90\\ 
Jun 10&\phantom{0}5.329&14.47&13.92\\ 
Jun 10&\phantom{0}5.356&14.44&13.88\\ 
Jun 10&\phantom{0}5.383&14.45&13.90\\ 
Jun 10&\phantom{0}5.410&14.48&13.93\\ 
Jun 10&\phantom{0}5.437&14.47&13.92\\ 
Jun 10&\phantom{0}5.464&14.45&13.89\\ 
Jun 10&\phantom{0}5.491&14.46&13.91\\ 
Jun 10&\phantom{0}5.519&14.47&13.91\\ 
Jun 10&\phantom{0}5.573&14.48&13.93\\ 
Jun 10&\phantom{0}5.600&14.48&13.94\\ 
Jun 10&\phantom{0}5.627&14.49&13.94\\ 
Jun 10&\phantom{0}5.990&14.48&13.94\\ 
Jun 10&\phantom{0}6.017&14.52&13.99\\ 
Jun 10&\phantom{0}6.043&14.50&13.96\\ 
Jun 10&\phantom{0}6.070&14.51&13.98\\ 
Jun 10&\phantom{0}6.097&14.52&13.99\\ 
Jun 10&\phantom{0}6.124&14.54&14.02\\ 
Jun 10&\phantom{0}6.151&14.55&14.03\\ 
Jun 10&\phantom{0}6.178&14.56&14.05\\ 
Jun 10&\phantom{0}6.204&14.54&14.02\\ 
Jun 10&\phantom{0}6.231&14.53&14.01\\ 
Jun 10&\phantom{0}6.460&14.62&14.13\\ 
Jun 10&\phantom{0}6.487&14.65&14.18\\ 
Jun 10&\phantom{0}6.513&14.64&14.16\\ 
Jun 10&\phantom{0}6.540&14.61&14.12\\ 
Jun 10&\phantom{0}6.567&14.63&14.14\\ 
Jun 10&\phantom{0}6.594&14.65&14.18\\ 
Jun 10&\phantom{0}6.621&14.67&14.20\\ 
Jun 10&\phantom{0}6.648&14.65&14.18\\ 
Jun 10&\phantom{0}6.674&14.66&14.19\\ 
Jun 10&\phantom{0}6.701&14.65&14.18\\ 
Jun 10&\phantom{0}6.921&14.71&14.26\\ 
Jun 10&\phantom{0}6.948&14.71&14.26\\ 
Jun 10&\phantom{0}6.974&14.71&14.26\\ 
Jun 10&\phantom{0}7.001&14.72&14.28\\ 
Jun 10&\phantom{0}7.028&14.71&14.27\\ 
Jun 10&\phantom{0}7.055&14.71&14.27\\ 
Jun 10&\phantom{0}7.082&14.72&14.28\\ 
Jun 10&\phantom{0}7.109&14.74&14.31\\ 
Jun 10&\phantom{0}7.136&14.74&14.31\\ 
Jun 10&\phantom{0}7.162&14.74&14.32\\ 
Jun 10&\phantom{0}7.231&14.73&14.29\\ 
Jun 10&\phantom{0}7.258&14.73&14.30\\ 
Jun 10&\phantom{0}7.285&14.73&14.29\\ 
Jun 10&\phantom{0}7.339&14.75&14.33\\ 
Jun 10&\phantom{0}7.366&14.73&14.29\\ 
Jun 10&\phantom{0}7.392&14.72&14.28\\ 
Jun 10&\phantom{0}7.446&14.70&14.26\\ 
Jun 10&\phantom{0}7.473&14.70&14.25\\ 
Jun 10&\phantom{0}7.505&14.69&14.24\\ 
Jun 10&\phantom{0}7.532&14.70&14.24\\ 
Jun 10&\phantom{0}7.559&14.70&14.25\\ 
Jun 10&\phantom{0}7.586&14.70&14.24\\ 
Jun 10&\phantom{0}7.613&14.68&14.22\\ 
Jun 10&\phantom{0}7.640&14.68&14.23\\ 
Jun 10&\phantom{0}7.667&14.69&14.23\\ 
Jun 10&\phantom{0}7.693&14.68&14.22\\ 
Jun 10&\phantom{0}7.720&14.67&14.21\\ 
Jun 10&\phantom{0}7.747&14.68&14.21\\ 
Jun 10&\phantom{0}8.012&14.64&14.16\\ 
Jun 10&\phantom{0}8.039&14.64&14.16\\ 
Jun 10&\phantom{0}8.066&14.63&14.14\\ 
Jun 10&\phantom{0}8.092&14.64&14.16\\ 
Jun 10&\phantom{0}8.119&14.63&14.14\\ 
Jun 10&\phantom{0}8.146&14.62&14.13\\ 
Jun 10&\phantom{0}8.227&14.61&14.12\\ 
Jun 10&\phantom{0}8.254&14.61&14.12\\ 
Jun 10&\phantom{0}9.289&14.46&13.91\\ 
Jun 10&\phantom{0}9.315&14.47&13.92\\ 
Jun 10&\phantom{0}9.342&14.45&13.90\\ 
Jun 10&\phantom{0}9.369&14.45&13.90\\ 
Jun 10&\phantom{0}9.396&14.46&13.91\\ 
Jun 10&\phantom{0}9.423&14.45&13.89\\ 
Jun 10&\phantom{0}9.450&14.45&13.90\\ 
Jun 10&\phantom{0}9.477&14.46&13.91\\ 
Jun 10&\phantom{0}9.503&14.44&13.88\\ 
Jun 10&\phantom{0}9.533&14.43&13.88\\ 
Jun 10&\phantom{0}9.560&14.43&13.87\\ 
Jun 10&\phantom{0}9.586&14.43&13.86\\ 
Jun 10&\phantom{0}9.613&14.43&13.87\\ 
Jun 10&\phantom{0}9.640&14.43&13.86\\ 
Jun 10&\phantom{0}9.667&14.42&13.86\\ 
Jun 10&\phantom{0}9.694&14.44&13.88\\ 
Jun 10&\phantom{0}9.721&14.47&13.92\\ 
Jun 10&\phantom{0}9.748&14.46&13.91\\ 
Jun 10&\phantom{0}9.775&14.46&13.91\\ 
Jun 10&10.024&14.43&13.86\\ 
Jun 10&10.051&14.43&13.87\\ 
Jun 10&10.078&14.43&13.86\\ 
Jun 10&10.105&14.43&13.88\\ 
Jun 10&10.132&14.43&13.87\\ 
Jun 10&10.159&14.45&13.90\\ 
Jun 10&10.185&14.44&13.89\\ 
Jun 10&10.212&14.45&13.90\\ 
Jun 10&10.239&14.44&13.89\\ 
Jun 10&10.266&14.43&13.87\\ 
Jun 10&10.496&14.45&13.90\\ 
Jun 10&10.523&14.49&13.95\\ 
Jun 10&10.550&14.50&13.97\\ 
Jun 10&10.577&14.51&13.98\\ 
Jun 10&10.604&14.49&13.95\\ 
Jun 10&10.769&14.51&13.98\\ 
Jun 10&10.796&14.51&13.98\\ 
Jun 10&10.823&14.54&14.02\\ 
Jun 10&10.850&14.52&13.99\\ 
Jun 10&10.877&14.55&14.03\\ 
Jun 10&11.001&14.57&14.06\\ 
Jun 10&11.028&14.56&14.05\\ 
Jun 10&11.055&14.59&14.09\\ 
Jun 10&11.180&14.60&14.11\\ 
Jun 10&11.207&14.60&14.10\\ 
Jun 10&11.234&14.62&14.14\\ 
Jun 10&11.263&14.62&14.13\\ 
\\
Jun 11&\phantom{0}4.516&14.54&14.14\\ 
Jun 11&\phantom{0}4.544&14.53&14.13\\ 
Jun 11&\phantom{0}4.573&14.50&14.09\\ 
Jun 11&\phantom{0}4.616&14.54&14.14\\ 
Jun 11&\phantom{0}4.644&14.60&14.23\\ 
Jun 11&\phantom{0}5.322&14.70&14.39\\ 
Jun 11&\phantom{0}5.349&14.69&14.38\\ 
Jun 11&\phantom{0}5.376&14.68&14.36\\ 
Jun 11&\phantom{0}5.403&14.68&14.36\\ 
Jun 11&\phantom{0}5.429&14.68&14.37\\ 
Jun 11&\phantom{0}5.459&14.68&14.36\\ 
Jun 11&\phantom{0}5.486&14.69&14.38\\ 
Jun 11&\phantom{0}5.513&14.73&14.43\\ 
Jun 11&\phantom{0}5.540&14.69&14.38\\ 
Jun 11&\phantom{0}5.567&14.67&14.35\\ 
Jun 11&\phantom{0}5.596&14.72&14.42\\ 
Jun 11&\phantom{0}5.623&14.69&14.37\\ 
Jun 11&\phantom{0}5.650&14.69&14.37\\ 
Jun 11&\phantom{0}5.677&14.69&14.37\\ 
Jun 11&\phantom{0}5.704&14.69&14.38\\ 
Jun 11&\phantom{0}5.976&14.65&14.31\\ 
Jun 11&\phantom{0}6.003&14.64&14.30\\ 
Jun 11&\phantom{0}6.030&14.64&14.29\\ 
Jun 11&\phantom{0}6.057&14.57&14.19\\ 
Jun 11&\phantom{0}6.084&14.58&14.20\\ 
Jun 11&\phantom{0}6.215&14.58&14.20\\ 
Jun 11&\phantom{0}6.242&14.57&14.19\\ 
Jun 11&\phantom{0}6.269&14.55&14.16\\ 
Jun 11&\phantom{0}6.296&14.55&14.15\\ 
Jun 11&\phantom{0}6.323&14.55&14.15\\ 
Jun 11&\phantom{0}6.487&14.50&14.08\\ 
Jun 11&\phantom{0}6.514&14.50&14.08\\ 
Jun 11&\phantom{0}6.541&14.49&14.06\\ 
Jun 11&\phantom{0}6.568&14.49&14.07\\ 
Jun 11&\phantom{0}6.594&14.49&14.07\\ 
Jun 11&\phantom{0}6.657&14.47&14.04\\ 
Jun 11&\phantom{0}6.684&14.47&14.04\\ 
Jun 11&\phantom{0}6.711&14.47&14.04\\ 
Jun 11&\phantom{0}6.737&14.48&14.06\\ 
Jun 11&\phantom{0}6.764&14.46&14.03\\ 
Jun 11&\phantom{0}6.906&14.45&14.01\\ 
Jun 11&\phantom{0}6.933&14.45&14.02\\ 
Jun 11&\phantom{0}6.960&14.46&14.03\\ 
Jun 11&\phantom{0}6.986&14.44&14.00\\ 
Jun 11&\phantom{0}7.013&14.44&13.99\\ 
Jun 11&\phantom{0}7.076&14.42&13.97\\ 
Jun 11&\phantom{0}7.138&14.41&13.95\\ 
Jun 11&\phantom{0}7.165&14.40&13.94\\ 
Jun 11&\phantom{0}7.192&14.39&13.93\\ 
Jun 11&\phantom{0}7.219&14.38&13.91\\ 
Jun 11&\phantom{0}7.246&14.37&13.90\\ 
Jun 11&\phantom{0}7.515&14.37&13.90\\ 
Jun 11&\phantom{0}7.541&14.37&13.90\\ 
Jun 11&\phantom{0}7.568&14.36&13.88\\ 
Jun 11&\phantom{0}7.595&14.37&13.89\\ 
Jun 11&\phantom{0}7.622&14.35&13.87\\ 
Jun 11&\phantom{0}7.651&14.36&13.88\\ 
Jun 11&\phantom{0}7.678&14.36&13.88\\ 
Jun 11&\phantom{0}7.705&14.35&13.87\\ 
Jun 11&\phantom{0}7.732&14.35&13.87\\ 
Jun 11&\phantom{0}7.759&14.34&13.86\\ 
Jun 11&\phantom{0}7.788&14.35&13.86\\ 
Jun 11&\phantom{0}7.815&14.35&13.87\\ 
Jun 11&\phantom{0}7.842&14.35&13.87\\ 
Jun 11&\phantom{0}7.869&14.36&13.88\\ 
Jun 11&\phantom{0}7.896&14.36&13.89\\ 
Jun 11&\phantom{0}8.093&14.38&13.91\\ 
Jun 11&\phantom{0}8.120&14.38&13.91\\ 
Jun 11&\phantom{0}8.147&14.38&13.90\\ 
Jun 11&\phantom{0}8.174&14.39&13.92\\ 
Jun 11&\phantom{0}8.201&14.39&13.92\\ 
Jun 11&\phantom{0}8.355&14.39&13.92\\ 
Jun 11&\phantom{0}8.382&14.39&13.93\\ 
Jun 11&\phantom{0}8.409&14.40&13.94\\ 
Jun 11&\phantom{0}8.436&14.40&13.94\\ 
Jun 11&\phantom{0}8.463&14.39&13.92\\ 
Jun 11&\phantom{0}8.689&14.42&13.96\\ 
Jun 11&\phantom{0}8.716&14.42&13.97\\ 
Jun 11&\phantom{0}8.743&14.43&13.97\\ 
Jun 11&\phantom{0}8.770&14.42&13.97\\ 
Jun 11&\phantom{0}8.797&14.43&13.98\\ 
Jun 11&\phantom{0}8.920&14.43&13.99\\ 
Jun 11&\phantom{0}8.947&14.43&13.99\\ 
Jun 11&\phantom{0}8.974&14.44&14.01\\ 
Jun 11&\phantom{0}9.001&14.45&14.01\\ 
Jun 11&\phantom{0}9.028&14.45&14.01\\ 
Jun 11&\phantom{0}9.158&14.47&14.04\\ 
Jun 11&\phantom{0}9.184&14.48&14.06\\ 
Jun 11&\phantom{0}9.211&14.48&14.06\\ 
Jun 11&\phantom{0}9.238&14.48&14.06\\ 
Jun 11&\phantom{0}9.265&14.49&14.07\\ 
Jun 11&\phantom{0}9.329&14.54&14.14\\ 
Jun 11&\phantom{0}9.396&14.51&14.09\\ 
Jun 11&\phantom{0}9.423&14.50&14.09\\ 
Jun 11&\phantom{0}9.449&14.51&14.11\\ 
Jun 11&\phantom{0}9.476&14.52&14.11\\ 
Jun 11&\phantom{0}9.503&14.52&14.11\\ 
\\
Jun 22&\phantom{0}7.910&14.15&13.97\\ 
Jun 22&\phantom{0}7.942&14.12&13.91\\ 
Jun 22&\phantom{0}7.971&14.11&13.91\\ 
Jun 22&\phantom{0}7.998&14.13&13.93\\ 
Jun 22&\phantom{0}8.027&14.11&13.89\\ 
Jun 22&\phantom{0}8.085&14.13&13.93\\ 
Jun 22&\phantom{0}8.143&14.13&13.93\\ 
Jun 22&\phantom{0}8.173&14.14&13.95\\ 
Jun 22&\phantom{0}8.200&14.14&13.95\\ 
Jun 22&\phantom{0}8.227&14.13&13.93\\ 
Jun 22&\phantom{0}8.368&14.15&13.96\\ 
Jun 22&\phantom{0}8.394&14.16&13.97\\ 
Jun 22&\phantom{0}8.421&14.15&13.96\\ 
Jun 22&\phantom{0}8.669&14.16&13.97\\ 
Jun 22&\phantom{0}8.695&14.16&13.98\\ 
Jun 22&\phantom{0}8.722&14.18&14.01\\ 
Jun 22&\phantom{0}8.946&14.23&14.10\\ 
Jun 22&\phantom{0}8.973&14.23&14.11\\ 
Jun 22&\phantom{0}9.000&14.25&14.13\\ 
Jun 22&\phantom{0}9.124&14.25&14.14\\ 
Jun 22&\phantom{0}9.151&14.27&14.17\\ 
Jun 22&\phantom{0}9.178&14.26&14.15\\ 
Jun 22&\phantom{0}9.489&14.30&14.23\\ 
Jun 22&\phantom{0}9.516&14.30&14.23\\ 
Jun 22&\phantom{0}9.543&14.30&14.24\\ 
Jun 22&\phantom{0}9.570&14.31&14.25\\ 
Jun 22&\phantom{0}9.597&14.31&14.25\\ 
Jun 22&\phantom{0}9.650&14.31&14.26\\ 
Jun 22&\phantom{0}9.707&14.33&14.29\\ 
Jun 22&\phantom{0}9.734&14.34&14.31\\ 
Jun 22&\phantom{0}9.761&14.33&14.30\\ 
Jun 22&\phantom{0}9.788&14.34&14.31\\ 
Jun 22&10.047&14.36&14.34\\ 
Jun 22&10.074&14.35&14.33\\ 
Jun 22&10.101&14.35&14.32\\ 
Jun 22&10.434&14.34&14.32\\ 
Jun 22&10.461&14.33&14.30\\ 
Jun 22&10.488&14.32&14.27\\ 
Jun 22&10.687&14.34&14.31\\ 
Jun 22&10.714&14.32&14.27\\ 
Jun 22&10.741&14.31&14.26\\ 
Jun 22&10.949&14.26&14.15\\ 
Jun 22&10.976&14.25&14.14\\ 
Jun 22&11.002&14.25&14.13\\ 
Jun 22&11.059&14.25&14.15\\ 
Jun 22&11.086&14.26&14.15\\ 
Jun 22&11.113&14.24&14.13\\ 
\\
Jun 23&\phantom{0}3.963&14.33&14.34\\ 
Jun 23&\phantom{0}3.999&14.34&14.36\\ 
Jun 23&\phantom{0}4.026&14.33&14.33\\ 
Jun 23&\phantom{0}4.052&14.34&14.35\\ 
Jun 23&\phantom{0}4.079&14.33&14.33\\ 
Jun 23&\phantom{0}4.106&14.35&14.37\\ 
Jun 23&\phantom{0}4.492&14.26&14.19\\ 
Jun 23&\phantom{0}4.519&14.24&14.17\\ 
Jun 23&\phantom{0}4.546&14.24&14.17\\ 
Jun 23&\phantom{0}4.606&14.23&14.15\\ 
Jun 23&\phantom{0}4.676&14.21&14.11\\ 
Jun 23&\phantom{0}4.703&14.21&14.11\\ 
Jun 23&\phantom{0}4.730&14.20&14.10\\ 
Jun 23&\phantom{0}4.963&14.17&14.04\\ 
Jun 23&\phantom{0}4.990&14.17&14.04\\ 
Jun 23&\phantom{0}5.017&14.18&14.05\\ 
Jun 23&\phantom{0}5.202&14.12&13.95\\ 
Jun 23&\phantom{0}5.229&14.14&13.97\\ 
Jun 23&\phantom{0}5.256&14.13&13.97\\ 
Jun 23&\phantom{0}5.444&14.11&13.93\\ 
Jun 23&\phantom{0}5.471&14.11&13.92\\ 
Jun 23&\phantom{0}5.498&14.10&13.92\\ 
Jun 23&\phantom{0}5.628&14.09&13.90\\ 
Jun 23&\phantom{0}5.655&14.10&13.90\\ 
Jun 23&\phantom{0}5.682&14.11&13.92\\ 
Jun 23&\phantom{0}5.710&14.11&13.92\\ 
Jun 23&\phantom{0}5.737&14.11&13.93\\ 
Jun 23&\phantom{0}5.764&14.10&13.92\\ 
Jun 23&\phantom{0}5.808&14.08&13.88\\ 
Jun 23&\phantom{0}5.834&14.07&13.87\\ 
Jun 23&\phantom{0}5.861&14.08&13.89\\ 
Jun 23&\phantom{0}5.894&14.07&13.87\\ 
Jun 23&\phantom{0}5.921&14.08&13.88\\ 
Jun 23&\phantom{0}5.947&14.09&13.89\\ 
Jun 23&\phantom{0}6.123&14.08&13.88\\ 
Jun 23&\phantom{0}6.150&14.09&13.90\\ 
Jun 23&\phantom{0}6.177&14.09&13.89\\ 
Jun 23&\phantom{0}6.341&14.09&13.89\\ 
Jun 23&\phantom{0}6.368&14.10&13.91\\ 
Jun 23&\phantom{0}6.395&14.11&13.93\\ 
Jun 23&\phantom{0}6.563&14.11&13.93\\ 
Jun 23&\phantom{0}6.590&14.13&13.96\\ 
Jun 23&\phantom{0}6.617&14.12&13.94\\ 
Jun 23&\phantom{0}6.773&14.13&13.96\\ 
Jun 23&\phantom{0}6.800&14.14&13.98\\ 
Jun 23&\phantom{0}6.827&14.14&13.99\\ 
Jun 23&\phantom{0}7.473&14.22&14.12\\ 
Jun 23&\phantom{0}7.500&14.23&14.15\\ 
Jun 23&\phantom{0}7.527&14.23&14.15\\ 
Jun 23&\phantom{0}7.557&14.24&14.17\\ 
Jun 23&\phantom{0}7.584&14.24&14.16\\ 
Jun 23&\phantom{0}7.611&14.24&14.17\\ 
Jun 23&\phantom{0}7.759&14.34&14.35\\ 
Jun 23&\phantom{0}7.786&14.32&14.32\\ 
Jun 23&\phantom{0}7.812&14.28&14.24\\ 
Jun 23&\phantom{0}7.843&14.28&14.24\\ 
Jun 23&\phantom{0}7.870&14.30&14.27\\ 
Jun 23&\phantom{0}7.897&14.30&14.28\\ 
Jun 23&\phantom{0}7.929&14.29&14.26\\ 
Jun 23&\phantom{0}7.956&14.29&14.26\\ 
Jun 23&\phantom{0}7.982&14.30&14.28\\ 
Jun 23&\phantom{0}8.012&14.30&14.29\\ 
Jun 23&\phantom{0}8.039&14.29&14.27\\ 
Jun 23&\phantom{0}8.066&14.30&14.27\\ 
Jun 23&\phantom{0}8.097&14.30&14.28\\ 
Jun 23&\phantom{0}8.124&14.30&14.27\\ 
Jun 23&\phantom{0}8.151&14.31&14.30\\ 
Jun 23&\phantom{0}8.315&14.28&14.24\\ 
Jun 23&\phantom{0}8.342&14.29&14.26\\ 
Jun 23&\phantom{0}8.368&14.30&14.27\\ 
Jun 23&\phantom{0}8.496&14.26&14.20\\ 
Jun 23&\phantom{0}8.523&14.26&14.21\\ 
Jun 23&\phantom{0}8.549&14.26&14.21\\ 
Jun 23&\phantom{0}8.761&14.24&14.16\\ 
Jun 23&\phantom{0}8.788&14.23&14.14\\ 
Jun 23&\phantom{0}8.815&14.24&14.16\\ 
Jun 23&\phantom{0}8.944&14.21&14.11\\ 
Jun 23&\phantom{0}8.971&14.21&14.11\\ 
Jun 23&\phantom{0}8.998&14.20&14.10\\ 
Jun 23&\phantom{0}9.149&14.20&14.08\\ 
Jun 23&\phantom{0}9.176&14.20&14.10\\ 
Jun 23&\phantom{0}9.203&14.20&14.09\\ 
Jun 23&\phantom{0}9.407&14.19&14.07\\ 
Jun 23&\phantom{0}9.434&14.18&14.05\\ 
Jun 23&\phantom{0}9.461&14.17&14.04\\ 
Jun 23&\phantom{0}9.607&14.17&14.03\\ 
Jun 23&\phantom{0}9.634&14.16&14.02\\ 
Jun 23&\phantom{0}9.661&14.17&14.03\\ 
Jun 23&\phantom{0}9.868&14.14&13.99\\ 
Jun 23&\phantom{0}9.895&14.13&13.97\\ 
Jun 23&\phantom{0}9.922&14.15&14.00\\ 
Jun 23&10.021&14.15&14.00\\ 
Jun 23&10.048&14.14&13.99\\ 
Jun 23&10.075&14.15&14.00\\ 
Jun 23&10.103&14.13&13.97\\ 
Jun 23&10.130&14.13&13.97\\ 
Jun 23&10.157&14.11&13.94\\ 
Jun 23&10.186&14.11&13.93\\ 
Jun 23&10.213&14.12&13.95\\ 
Jun 23&10.240&14.12&13.94\\ 
Jun 23&10.269&14.12&13.95\\ 
Jun 23&10.296&14.09&13.90\\ 
Jun 23&10.323&14.09&13.90\\ 
Jun 23&10.357&14.11&13.92\\ 
Jun 23&10.384&14.09&13.89\\ 
Jun 23&10.411&14.10&13.91\\ 
Jun 23&10.524&14.11&13.92\\ 
Jun 23&10.551&14.11&13.92\\ 
Jun 23&10.578&14.13&13.96\\ 
Jun 23&10.709&14.11&13.92\\ 
Jun 23&10.742&14.12&13.94\\ 
\hline
\hline
\footnotetext[1]{UT at midpoint of the exposure (uncorrected for light travel time).}
\footnotetext[2]{Observed R-band magnitude (after applying absolute calibrations, extinction corrections, and comparison star corrections) in a circular aperture with radius 6 arcsec.}
\footnotetext[3]{Coma-removed m$_R$(1,1,0) corrected by $\Delta$m$_2$ (given in Table~\ref{t:obs_summary}) so that all nights have the same peak magnitude.}
\end{longtable}
\end{center}
\renewcommand{\thefootnote}{\arabic{footnote}}


\clearpage

\renewcommand{\thefootnote}{\alph{footnote}}
\scriptsize
\begin{center}
\begin{longtable}{lcccccccccccc}
\caption[Table of photometry]{Table of photometry}\label{t:obs_summary_1988}\\
\hline
\hline\\[-1.0ex]
  \multicolumn{1}{l}{UT Date\,\tablenotemark{a}}&
  \multicolumn{1}{c}{Tel.\,\tablenotemark{b}}&
  \multicolumn{1}{c}{Instrument}&
  \multicolumn{1}{c}{Primary}&
  \multicolumn{1}{c}{$r$}&
  \multicolumn{1}{c}{$\Delta$}&
  \multicolumn{1}{c}{$\alpha$}&
  \multicolumn{1}{c}{$\Delta$t}&
  \multicolumn{1}{c}{Ecl. Long.}&
  \multicolumn{1}{c}{Ecl. Long.}&
  \multicolumn{1}{c}{$\Delta$m$_1$\,\tablenotemark{f}}&
  \multicolumn{1}{c}{$\Delta$m$_2$\,\tablenotemark{g}}&
  \multicolumn{1}{c}{Ref.\,\tablenotemark{h}}\\
  \multicolumn{1}{l}{}&
  \multicolumn{1}{c}{}&
  \multicolumn{1}{c}{}&
  \multicolumn{1}{c}{Filter}&
  \multicolumn{1}{c}{(AU)}&
  \multicolumn{1}{c}{(AU)}&
  \multicolumn{1}{c}{($^\circ$)}&
  \multicolumn{1}{c}{(hr)\,\tablenotemark{c}}&
  \multicolumn{1}{c}{Earth ($^\circ$)\,\tablenotemark{d}}&
  \multicolumn{1}{c}{Sun ($^\circ$)\,\tablenotemark{e}}&
  \multicolumn{1}{c}{}&
  \multicolumn{1}{c}{}&
  \multicolumn{1}{c}{}\\[0.8ex]
  \hline\\[-1ex]
\endfirsthead

\multicolumn{13}{c}%
{{\bfseries \tablename\ \thetable{} -- continued from previous page}} \\[0.8ex]
\hline\\[-1.8ex]
\endhead

\\[-3ex]
\hline
\hline\\[-1.8ex]
 \multicolumn{13}{c}{{Footnotes continue on next page}} \\
\endfoot

\endlastfoot
1987 Mar 31&KP2.1&TI 2&R&3.991&3.162&\phantom{0}9.0&0.438&332.6&341.6&$-$5.79&\phantom{0}$+$0.45&JM\\
1987 Apr 01&KP2.1&TI 2&R&3.987&3.168&\phantom{0}9.2&0.439&332.5&341.7&$-$5.80&\phantom{0}$+$0.45&JM\\
1987 Apr 02&KP2.1&TI 2&R&3.984&3.174&\phantom{0}9.4&0.440&332.3&341.8&$-$5.81&\phantom{0}$+$0.45&JM\\
1987 Apr 03&KP2.1&TI 2&R&3.981&3.181&\phantom{0}9.7&0.441&332.2&341.9&$-$5.82&\phantom{0}$+$0.45&JM\\
1988 Feb 25&MH2.4&MASCOT&R&2.384&1.985&24.0&0.275&\phantom{0}54.3&\phantom{0}29.9&$-$4.14&\phantom{0}$-$0.05&JL\\
1988 Feb 27&MH2.4&MASCOT&R&2.371&1.949&24.1&0.270&\phantom{0}54.9&\phantom{0}30.4&$-$4.09&\phantom{0}$-$0.05&JL\\
1988 Feb 28&MH2.4&MASCOT&R&2.365&1.932&24.1&0.268&\phantom{0}55.1&\phantom{0}30.7&$-$4.07&\phantom{0}$-$0.05&JL\\
1988 Feb 29&MH2.4&MASCOT&R&2.359&1.914&24.1&0.265&\phantom{0}55.4&\phantom{0}30.9&$-$4.04&\phantom{0}$-$0.05&JL\\
1988 Apr 09&MH1.3&MASCOT&R&2.103&1.279&20.0&0.177&\phantom{0}61.7&\phantom{0}42.8&$-$2.79&\phantom{0}$+$0.30&JL\\
1988 Apr 10&MH1.3&MASCOT&R&2.097&1.265&19.8&0.175&\phantom{0}61.7&\phantom{0}43.1&$-$2.75&\phantom{0}$+$0.27&JL\\
1988 Apr 12&MH1.3&MASCOT&R&2.084&1.239&19.3&0.172&\phantom{0}61.8&\phantom{0}43.8&$-$2.68&\phantom{0}$+$0.30&JL\\
1988 Apr 15&MH1.3&MASCOT&R&2.065&1.201&18.6&0.166&\phantom{0}61.8&\phantom{0}44.8&$-$2.57&\phantom{0}$+$0.30&JL\\[-0.4ex]
1988 May 08&SO2.3&TI&VR\tablenotemark{i}&1.919&0.959&13.2&0.133&\phantom{0}59.3&\phantom{0}53.4&$-$1.75&...&WI\\
1988 May 09&SO2.3&TI&VR&1.913&0.951&13.1&0.132&\phantom{0}59.1&\phantom{0}53.8&$-$1.72&...&WI\\
1988 May 20&ML1.5&photometer&VR&1.845&0.873&12.8&0.121&\phantom{0}56.5&\phantom{0}58.3&$-$1.44&\phantom{0}$-$0.10&WI\\
1988 May 21&ML1.5&photometer&VR&1.839&0.867&12.9&0.120&\phantom{0}56.2&\phantom{0}58.7&$-$1.43&\phantom{0}$-$0.10&WI\\
1988 May 22&ML1.5&photometer&VR&1.832&0.861&13.1&0.119&\phantom{0}56.0&\phantom{0}59.2&$-$1.41&\phantom{0}$-$0.10&WI\\
1988 Jun 10&IRTF&bolometer&N&1.720&0.788&19.9&0.109&\phantom{0}51.3&\phantom{0}68.0&$-$1.30&$+$11.80&AH\\[-0.4ex]
1988 Jun 10&UH2.2&photometer&6840\AA\tablenotemark{j}&1.720&0.788&19.9&0.109&\phantom{0}51.3&\phantom{0}68.0&$-$1.30&\phantom{0}$+$0.73&AH\\
1988 Jun 11&IRTF&bolometer&N&1.714&0.785&20.4&0.109&\phantom{0}51.1&\phantom{0}68.4&$-$1.30&$+$11.70&AH\\[-0.4ex]
1988 Jun 11&UH2.2&photometer&6840\AA\tablenotemark{j}&1.714&0.785&20.4&0.109&\phantom{0}51.1&\phantom{0}68.4&$-$1.30&$\phantom{0}+$0.73&AH\\
1988 Jun 12&ML1.5&photometer&VR&1.708&0.783&20.9&0.108&\phantom{0}50.9&\phantom{0}68.9&$-$1.30&\phantom{0}$+$0.25&WI\\
1988 Jun 14&ML1.5&photometer&VR&1.697&0.780&21.9&0.108&\phantom{0}50.5&\phantom{0}69.9&$-$1.31&\phantom{0}$+$0.25&WI\\
1988 Jun 22&MH2.4&BRICC&R&1.653&0.771&25.9&0.107&\phantom{0}49.5&\phantom{0}74.1&$-$1.36&\phantom{0}$+$0.23&JL\\
1988 Jun 23&MH2.4&BRICC&R&1.647&0.770&26.3&0.107&\phantom{0}49.4&\phantom{0}74.6&$-$1.36&\phantom{0}$+$0.28&JL\\
1988 Jun 30&MH2.4&BRICC&R&1.611&0.769&29.7&0.107&\phantom{0}49.2&\phantom{0}78.4&$-$1.42&\phantom{0}$+$0.45&JL\\
1994 Oct 26&KP0.9&T2KA&R&2.498&1.640&14.2&0.227&247.5&234.2&$-$3.52&...&MF\\
1994 Dec 18&KP0.9&T2KA&R&2.802&1.932&11.3&0.268&234.8&245.5&$-$4.03&...&MF\\
1994 Dec 19&KP0.9&T2KA&R&2.807&1.944&11.6&0.269&234.6&245.7&$-$4.06&...&MF\\
\\[-0.07in]
\hline
\hline
\footnotetext[1]{Parameters were determined for 7:00 UT on the date of observation.}\\[-0.1in]
\footnotetext[2]{Telescope abbreviations: IRTF = NASA Infrared Telescope, KP0.9 = Kitt Peak 0.9-m, MH1.3 = McGraw-Hill 1.3-m, MH2.4 = McGraw-Hill 2.4-m, ML1.5 = Mt. Lemmon 1.5-m, SO2.3 = Steward Observatory 2.3-m, UH2.2 = University of Hawaii 2.2-m.}
\footnotetext[3]{Light travel time.}
\footnotetext[4]{Ecliptic longitude of the Earth as seen from the comet.}
\footnotetext[5]{Ecliptic longitude of the Sun as seen from the comet.}
\footnotetext[6]{m$_R$(1,1,0) correction for the night (in magnitudes).}
\footnotetext[7]{Offset (in magnitudes) necessary to make data on all nights peak at the same magnitude.}
\footnotetext[8]{Reference abbreviations: AH = \citep{ahearn89}, JL = \citep{jewitt89}, JM = \citep{jewitt88}, MF = \citep{mueller96}, WI = \citep{wisniewski90}.}
\footnotetext[9]{The VR filter was centered at 5886 {\AA} with a 2160 {\AA} half-width \citep{wisniewski90}.}
\footnotetext[10]{A similar number of observations was also made at 4845 {\AA}. However, we primarily used the 6840 {\AA} data and so $\Delta$m$_2$ is only listed for the 6840 {\AA} data. Both of these filters are part of the International Halley Watch filter set \citep{osborn90}.}
\end{longtable}
\end{center}
\renewcommand{\thefootnote}{\arabic{footnote}}

\clearpage


\end{document}